\documentclass[letterpaper,titlepage,11pt]{article}
\usepackage{hyperref}

\usepackage{amssymb,amsmath,amsfonts}
\usepackage{epsfig}

\def\clock{{\count0=\time
           \divide\count0 60
           \ifnum\count0<10 0\fi\the\count0
           \multiply\count0 -60 \advance\count0 \time
           :\ifnum\count0<10 0\fi \the\count0
         }}
\newcommand{\timestamp}{{\small\vbox{\hbox{\tt\jobname.tex}
\hbox{\the\day/\the\month/\the\year, \clock}}}}

\newcommand{\CE}{\mathcal{E}}

\newcommand{\CH}{\mathcal{H}}

\newcommand{\CO}{\mathcal{O}}

\newcommand{\CM}{\mathcal{M}}
\newcommand{\CN}{\mathcal{N}}

\newcommand{\Z}{\mathbb{Z}}

\newcommand{\R}{\mathbb{R}}

\newcommand{\spa}{\ , \ \ }

\newcommand{\ds}{\displaystyle}

\newcommand{\tr}{\mathop{{\rm Tr}}}

\newcommand{\ads}{\mbox{AdS}}
\newcommand{\gym}{g_{\rm YM}}
\newcommand{\zeth}{\zeta \Big( \frac{3}{2} \Big)}

\numberwithin{equation}{section}

\begin{document}

\begin{titlepage}

\rightline{\vbox{\small\hbox{\tt hep-th/0608115} }}
 \vskip 2.7cm

\centerline{\LARGE \bf Matching the Hagedorn temperature in
AdS/CFT} \vskip 2cm

\centerline{\large {\bf  Troels Harmark} and {\bf Marta Orselli} }
\vskip 1.1cm
\begin{center}
\sl The Niels Bohr Institute and Nordita\\
\sl Blegdamsvej 17, 2100 Copenhagen \O , Denmark
\end{center}

\vskip 0.7cm

\centerline{\small\tt harmark@nbi.dk, orselli@nbi.dk}

\vskip 1.7cm

\centerline{\bf Abstract} \vskip 0.2cm \noindent We match the
Hagedorn/deconfinement temperature of planar $\CN=4$ super
Yang-Mills (SYM) on $\R \times S^3$ to the Hagedorn temperature of
string theory on $\ads_5 \times S^5$. The match is done in a
near-critical region where both gauge theory and string theory are
weakly coupled. The near-critical region is near a point with zero
temperature and critical chemical potential. On the gauge theory
side we are taking a decoupling limit found in hep-th/0605234 in
which the physics of planar $\CN=4$ SYM is given exactly by the
ferromagnetic $XXX_{1/2}$ Heisenberg spin chain. We find moreover
a general relation between the Hagedorn/deconfinement temperature
and the thermodynamics of the Heisenberg spin chain and we use
this to compute it in two distinct regimes. On the string theory
side, we identify the dual limit for which the string tension and
string coupling go to zero. This limit is taken of string theory
on a maximally supersymmetric pp-wave background with a flat
direction, obtained from a Penrose limit of $\ads_5 \times S^5$.
We compute the Hagedorn temperature of the string theory and find
agreement with the Hagedorn/deconfinement temperature computed on
the gauge theory side.


\end{titlepage}

\pagestyle{plain} \setcounter{page}{1}

\tableofcontents

\section{Introduction and summary}

The AdS/CFT correspondence states that $SU(N)$ $\CN = 4$ super
Yang-Mills (SYM) on $\R \times S^3$ is equivalent to string theory
on $\ads_5 \times S^5$
\cite{Maldacena:1997re,Gubser:1998bc,Witten:1998qj}. In
particular, planar $\CN = 4$ SYM on $\R \times S^3$ with 't Hooft
coupling $\lambda$ is conjectured to be equivalent to weakly
coupled string theory on $\ads_5 \times S^5$ with string
tension $T_{\rm str}$, with the relation%
\footnote{See main text for our conventions regarding $T_{\rm
str}$ and $\lambda$.}
\begin{equation}
T_{\rm str} = \frac{1}{2} \sqrt{\lambda}
\end{equation}
The most impressive checks on this correspondence have involved
computing physical quantities on the gauge theory side, such as
the expectation value of Wilson loops
\cite{Erickson:2000af,Drukker:2000rr} or the anomalous dimensions
of gauge theory operators \cite{Berenstein:2002jq}, and
extrapolating the results to strong coupling in order to compare
with string theory.

In this paper we take a different route. We compute the
Hagedorn/deconfinement temperature for planar $\CN = 4$ SYM on $\R
\times S^3$ at weak coupling $\lambda \ll 1$ in a certain
near-critical region found in \cite{Harmark:2006di}. We match then
this to the Hagedorn temperature computed in weakly coupled string
theory on $\ads_5 \times S^5$, in the corresponding dual
near-critical region. Beyond this, we successfully match the low
energy spectra of the gauge theory and the string theory in the
near-critical region. The matching of the spectra and Hagedorn
temperature is possible since we take a zero string tension limit
on the string theory side.

That the Hagedorn/deconfinement temperature of planar $\CN = 4$
SYM on $\R \times S^3$ is dual to the Hagedorn temperature of
string theory on $\ads_5 \times S^5$ was conjectured in
\cite{Witten:1998zw,Sundborg:1999ue,Polyakov:2001af,Aharony:2003sx}.
This originated in the finding of a confinement/deconfinement
phase transition in planar $\CN = 4$ SYM on $\R \times S^3$ at
weak coupling $\lambda \ll 1$ \cite{Witten:1998zw}. For large
energies the theory has a Hagedorn density of states, with the
Hagedorn temperature being equal to the deconfinement temperature
\cite{Sundborg:1999ue,Polyakov:2001af,Aharony:2003sx}.

On the string theory side, it is unfortunately not possible to
compute the Hagedorn temperature for string theory on
$\ads_5\times S^5$ since we do not know how to make a first
quantization of string theory in this background. However, in
certain Penrose limits, where the $\ads_5 \times S^5$ background
becomes a maximally supersymmetric pp-wave background
\cite{Berenstein:2002jq,Bertolini:2002nr}, one can find the string
spectrum, and the computation of the Hagedorn temperature has been
done
\cite{PandoZayas:2002hh,Greene:2002cd,Sugawara:2002rs,Brower:2002zx,
Sugawara:2003qc,Grignani:2003cs,Hyun:2003ks,Bigazzi:2003jk}.

From these facts it is clear that any successful matching of the
Hagedorn/deconfinement temperature for the gauge theory with the
Hagedorn temperature of string theory should be to the Hagedorn
temperature of the maximally supersymmetric pp-wave background.
Therefore, one should make the match for large R-charges/angular
momenta.

However, if we consider the pp-wave/gauge-theory correspondence of
\cite{Berenstein:2002jq} we encounter a problem. In
\cite{Berenstein:2002jq} the gauge theory states that are
conjectured to correspond to string states on the pp-wave side are
only a small subset of the possible gauge theory states. But, at
weak coupling $\lambda \ll 1$, all of these possible gauge theory
states are present. The crucial step of \cite{Berenstein:2002jq},
in order to resolve this problem, is to consider a strong coupling
limit $\lambda\rightarrow \infty$ on the gauge theory side in
which it is conjectured that most of the gauge theory states will
decouple, and only a small subset of the states, believed to be
precisely the ones dual to the string states, should remain in
this limit. More specifically, the ground state and zero modes of
the pp-wave string theory are mapped to chiral primary states in
$\CN =4$ SYM, and the surviving states in the large $\lambda$
limit should then be the states that lie sufficiently close to the
chiral primary states with respect to their anomalous dimensions.
Thus, seemingly, we cannot match the Hagedorn/deconfinement
temperature at weak coupling $\lambda \ll 1$ to the pp-wave
Hagedorn temperature, since on the gauge theory side we have many
more states than the ones dual to the pp-wave string states.

In this paper we resolve this problem by employing a recently
found decoupling limit of thermal $SU(N)$ $\CN=4$ SYM on $\R
\times S^3$ \cite{Harmark:2006di}. Denoting the three R-charges
for the $SU(4)$ R-symmetry as $J_i$, $i=1,2,3$, and their
corresponding chemical potentials as $\Omega_i$, $i=1,2,3$, and
putting $(\Omega_1,\Omega_2,\Omega_3)=(\Omega,\Omega,0)$, we can
write the decoupling limit as \cite{Harmark:2006di}
\begin{equation}
\label{ourlim} T\rightarrow 0\spa \Omega\rightarrow 1\spa
\lambda\rightarrow 0\spa \tilde{T}\equiv\frac{T}{1-\Omega}\
\mbox{fixed}\spa \tilde{\lambda}\equiv\frac{\lambda}{1-\Omega}\
\mbox{fixed} \spa N \ \mbox{fixed}
\end{equation}
where $T$ is the temperature for $\CN=4$ SYM. In this limit only
the states in the $SU(2)$ sector survive, and $SU(N)$ $\CN=4$ SYM
on $\R \times S^3$ reduces to a quantum mechanical theory with
temperature $\tilde{T}$ and coupling $\tilde{\lambda}$. In the
planar limit $N=\infty$, we have furthermore that in the limit
\eqref{ourlim} $\CN=4$ SYM on $\R \times S^3$ has the Hamiltonian
$D_0 + \tilde{\lambda}D_2$, where $D_0$ is the bare scaling
dimension and $D_2$ is the Hamiltonian for the ferromagnetic
$XXX_{1/2}$ Heisenberg spin chain (without magnetic field). We see
that the limit \eqref{ourlim} includes taking a zero 't Hooft
coupling limit $\lambda\rightarrow 0$, thus we are in weakly
coupled $\CN=4$ SYM after the limit.

The resolution to the above stated problem that there are too many
states for $\lambda \ll 1$ is now as follows. Since
$\tilde{\lambda}$ can be finite even though $\lambda \rightarrow 0$
we can consider in particular the $\tilde{\lambda} \gg 1$ region. In
this region the low energy states for the $D_2$ Hamiltonian are the
dominant states. These states are the vacua, plus the magnon states
of the Heisenberg spin chain. The vacua precisely consist of the
chiral primary sector of the $SU(2)$ sector. Therefore, by
considering $\tilde{\lambda} \gg 1$ we can circumvent the apparent
problem with matching the pp-wave spectrum to the spectrum of weakly
coupled gauge theory.

For planar $\CN=4$ SYM on $\R\times S^3$ in the decoupling limit
\eqref{ourlim} we find a direct connection between the
Hagedorn/deconfinement temperature for finite $\tilde{\lambda}$
and the thermodynamics of the Heisenberg spin chain. If we denote
$t$ as the temperature and $-tV(t)$ as the thermodynamic limit of
the free energy per site for the ferromagnetic Heisenberg chain
with Hamiltonian $D_2$, then the Hagedorn temperature
$\tilde{T}=\tilde{T}_H$ is given by
\begin{equation}
\tilde{T}_H = \frac{1}{V\!
\big(\tilde{\lambda}^{-1}\tilde{T}_H\big)}
\end{equation}
We use this to compute the Hagedorn temperature for small
$\tilde{\lambda}$, in which case it corresponds to the high
temperature limit of the Heisenberg chain. For large
$\tilde{\lambda}$ the Hagedorn temperature is instead mapped to the
low temperature limit of the Heisenberg chain, and we obtain in this
limit the Hagedorn temperature
\begin{equation}
\label{ourhag} \tilde{T}_H = (2\pi)^{\frac{1}{3}} \left[ \zeta
\Big( \frac{3}{2} \Big) \right]^{-\frac{2}{3} }
\tilde{\lambda}^{\frac{1}{3}}
\end{equation}
where $\zeta(x)$ is the Riemann zeta function. Note that we have
that the low energy behavior of the Heisenberg chain is tied to
the large $\tilde{\lambda}$ limit, as we also stated above. In
fact, the low energy spectrum consisting of the chiral primary
vacua with the magnon spectrum gives rise to the Hagedorn
temperature \eqref{ourhag}.

On the string theory side, we find using the AdS/CFT duality the
following decoupling limit of string theory on $\ads_5\times S^5$,
dual to the limit \eqref{ourlim},
\begin{equation}
\label{ouradslim} \epsilon \rightarrow 0 \spa \tilde{H} \equiv
\frac{E-J}{\epsilon} \ \mbox{fixed} \spa \tilde{T}_{\rm str}
\equiv \frac{T_{\rm str}}{\sqrt{\epsilon}} \ \mbox{fixed} \spa
\tilde{g}_s \equiv \frac{g_s}{\epsilon}\ \mbox{fixed} \spa J_i\
\mbox{fixed}
\end{equation}
Here $E$ is the energy of the strings, $J_i$, $i=1,2,3$, are the
angular momenta for the five-sphere, $J = J_1+J_2$, and $g_s$ is the
string coupling. $\tilde{H}$ is the effective Hamiltonian for the
strings in the decoupling limit. We see that both the string tension
$T_{\rm str}$ and the string coupling $g_s$ go to zero in this
limit.

The next step is to consider a Penrose limit of the $\ads_5\times
S^5$ background, and to consider the string theory on the
resulting pp-wave background. We note that the Penrose limit of
\cite{Berenstein:2002jq} does not result in the right light-cone
quantized string theory spectrum for our purposes. We need a
pp-wave spectrum for which all states with $E=J$, $J=J_1+J_2$,
correspond to the string vacua. This is precisely what the Penrose
limit of \cite{Bertolini:2002nr} provides. In more detail, on the
gauge theory/spin chain side, $J_1-J_2$ measures the total spin,
and we have a vacuum for each value of the total spin. The dual
manifestation of this is that in the pp-wave background that is
obtained using the Penrose limit of \cite{Bertolini:2002nr} we
have a flat direction, such that there is a vacuum for each value
of the momentum along that direction, and that momentum is
moreover dual to $J_1-J_2$.

We implement then the decoupling limit \eqref{ouradslim} for the
pp-wave background. This corresponds to a large $\mu$ limit of the
pp-wave, with $\mu$ being a parameter in front of the square well
potential terms for six of the eight bosonic directions. We show
that we get the same spectrum as that of the spectrum for large
$\tilde{\lambda}$ and $J$ of planar $\CN=4$ SYM on $\R\times S^3$
in the decoupling limit \eqref{ourlim}. Thus, we can match the
spectrum of weakly coupled string theory with weakly coupled gauge
theory in the decoupling limits.

We proceed to compute the Hagedorn temperature for string theory
on the pp-wave background in the large $\mu$ limit in two
different ways. The first way is to compute the Hagedorn
temperature from the spectrum obtained by taking the large $\mu$
limit directly on the spectrum. The second way is to take the
Hagedorn temperature for the full pp-wave spectrum, which was
computed in \cite{Sugawara:2003qc}, and take the large $\mu$ limit
of that. The two ways of computing the Hagedorn temperature agree,
which is a good check on the fact that most of the string states
really do decouple in the large $\mu$ limit. Moreover, the
resulting Hagedorn temperature can, via the AdS/CFT duality, be
compared to the Hagedorn/deconfinement temperature \eqref{ourhag}
computed in weakly coupled gauge theory, and they are shown to
agree.

In summary, we match the Hagedorn/deconfinement temperature
computed in weakly coupled planar $\CN=4$ SYM on $\R \times S^3$,
in the decoupling limit \eqref{ourlim}, to the Hagedorn
temperature computed on a maximally supersymmetric pp-wave
background in the dual decoupling limit \eqref{ouradslim}. The
fact that we are in a pp-wave background corresponds to being in
the large $J$ sector of string theory on $\ads_5 \times S^5$.
Moreover, we show that the low energy spectra of gauge theory and
string theory in the decoupling limit are the same, which can be
seen as the underlying reason for the matching of the Hagedorn
temperature. In the Conclusions in Section \ref{sec:concl} we
discuss the matching in the larger framework of a decoupled sector
of the AdS/CFT correspondence for which we have a spin chain/gauge
theory/string theory triality.

\section{The $SU(2)$ decoupling limit of $\CN=4$ SYM on $\R \times
S^3$} \label{sec:declim}

In this section we review the decoupling limit of $\CN=4$ SYM on
$\R \times S^3$ found in \cite{Harmark:2006di} in which $\CN=4$
SYM reduces to a quantum mechanical theory on the $SU(2)$ sector
which becomes the ferromagnetic $XXX_{1/2}$ Heisenberg spin chain
in the planar limit.

\subsubsection*{Thermal $\CN=4$ SYM on $\R \times S^3$ and the Hagedorn temperature}

In \cite{Harmark:2006di} the thermal partition function of $SU(N)$
$\CN=4$ SYM on $\R \times S^3$ with non-zero chemical potentials
is considered.%
\footnote{To be precise, it is in fact the gauge group $U(N)$
which is considered in \cite{Harmark:2006di}. In this paper we
consider instead the gauge group $SU(N)$. However, all the results
of \cite{Harmark:2006di} can easily be formulated for $SU(N)$
instead of $U(N)$. In particular, many of the considerations of
\cite{Harmark:2006di} concerns the planar limit $N=\infty$ in
which case the results for $SU(N)$ and $U(N)$ are equivalent.}
 We can write this in general as follows. Let $D$
denote the dilatation operator giving the scaling dimension for a
given operator (or energy of the corresponding state). Let $J_i$,
$i=1,2,3$, denote the three R-charges associated with the $SU(4)$
R-symmetry of $\CN=4$ SYM, and let $\Omega_i$ be the three
chemical potentials corresponding to these charges. Then we can
write the full partition function as
\begin{equation}
\label{genZ} Z(\beta,\Omega_i) = \tr \left( e^{-\beta D + \beta
\sum_{i=1}^3 \Omega_i J_i } \right)
\end{equation}
where $\beta=1/T$ is the inverse temperature. Here the trace is
taken over all gauge invariant states, corresponding to all the
multi-trace operators. When $\CN=4$ SYM is weakly coupled, we can
expand the dilatation operator in powers of the 't Hooft coupling
as follows \cite{Beisert:2003tq,Beisert:2004ry}
\begin{equation}
\label{genD} D = D_0 + \sum_{n=2}^\infty \lambda^{n/2} D_n
\end{equation}
where we have defined for our convenience the 't Hooft coupling as
\begin{equation}
\lambda = \frac{\gym^2 N}{4\pi^2}
\end{equation}
$\gym$ being the Yang-Mills coupling of $\CN=4$ SYM.

For free $SU(N)$ $\CN=4$ SYM on $\R \times S^3$ in the planar
limit $N=\infty$ the partition function exhibits a singularity at
a certain temperature $T_H$
\cite{Sundborg:1999ue,Polyakov:2001af,Aharony:2003sx}. The
temperature $T_H$ is a Hagedorn temperature for planar $\CN=4$ SYM
on $\R \times S^3$ since the density of states goes like
$e^{E/T_H}$ for high energies $E\gg 1$ (we work in units with
radius of the $S^3$ set to one). Moreover, we have that for $T <
T_H$ the partition function is of order one, while for $T > T_H$
the partition function is of order $N^2$, and for large
temperatures the partition function is like for free $SU(N)$
$\CN=4$ SYM on $\R^4$. Therefore we have a transition at $T_H$
resembling the confinement/deconfinement phase transition in QCD,
thus in this sense we can regard $T_H$ as a deconfinement
temperature for
planar $\CN=4$ SYM on $\R \times S^3$.%
\footnote{In \cite{Aharony:2005bq} it was found for weakly coupled
large $N$ pure Yang-Mills theory on $\R \times S^3$ that the
deconfinement temperature is lower than the Hagedorn temperature,
which means that this theory has a first order phase transition at
the deconfinement temperature.}

Turning on the coupling $\lambda$ and the chemical potentials
$\Omega_i$ the Hagedorn singularity for planar $\CN=4$ SYM on $\R
\times S^3$ persists, at least for $\lambda \ll 1$
\cite{Spradlin:2004pp,Yamada:2006rx,Harmark:2006di}. The Hagedorn
temperature $T_{\rm H}$ is a function of $\lambda$ and $\Omega_i$,
and it is known in certain limits. The first order correction in
$\lambda$ for $\Omega_i=0$ was found in \cite{Spradlin:2004pp}.
For $\lambda=0$ and non-zero chemical potentials $\Omega_i$ the
Hagedorn temperature was found in
\cite{Yamada:2006rx,Harmark:2006di} while the one-loop correction
was found in \cite{Harmark:2006di}. E.g. for weak coupling and
small chemical potentials it is found that \cite{Harmark:2006di}
\begin{equation}
T_H = \frac{1}{\beta_0} \left( 1 + \frac{\lambda}{2} \right) -
\frac{1}{6\sqrt{3}} \left( 1 - \frac{\lambda}{2}
(11-\beta_0\sqrt{3}) \right) \sum_{i=1}^3 \Omega_i^2 +
\CO(\lambda^2) + \CO(\Omega_i^4)
\end{equation}
with $\beta_0 \equiv -\log (7-4\sqrt{3})$. See
\cite{Harmark:2006di} for the fourth order correction in the
chemical potentials.

\subsubsection*{The $SU(2)$ decoupling limit}

It was found in \cite{Harmark:2006di} that near the critical point
$(T,\Omega_1,\Omega_2,\Omega_3)=(0,1,1,0)$ most of the states of
$\CN=4$ SYM decouple and we end up with a much simpler theory that
we can regard as quantum mechanical. In order to write the
decoupling limit we define the charge $J = J_1+J_2$ and we define
$\Omega$ as the corresponding chemical potential. In the following
we are interested in the situation for which
$\Omega_1=\Omega_2=\Omega$. We consider then the decoupling limit
\cite{Harmark:2006di}
\begin{equation}
\label{su2limit}
T \rightarrow 0 \spa \Omega \rightarrow 1 \spa
\lambda \rightarrow 0 \spa \tilde{T} \equiv \frac{T}{1-\Omega} \
\mbox{fixed} \spa \tilde{\lambda} \equiv \frac{\lambda}{1-\Omega}
\ \mbox{fixed}
\end{equation}
In this limit most of the states of $\CN=4$ SYM decouple. This is
due to the fact that only the states with $D-J$ being of order
$1-\Omega$ survive. Therefore the states that survive are the ones
with $D_0=J$, i.e. with the bare scaling dimension equal to $J$.
From this one can see that the total Hilbert space of the theory
consists of all states corresponding to all the multi-trace
operators that one can write down from the two complex scalars $Z$
and $X$, where $Z$ and $X$ have the R-charge weights $(1,0,0)$ and
$(0,1,0)$, respectively. Thus, we have that our Hilbert space
$\CH$ consists of all possible linear combinations of the
multi-trace operators%
\footnote{Here we will loosely refer to the single-trace or
multi-trace operators as states in a Hilbert space, the precise
meaning being that any single-trace or multi-trace operator $\CO$
for $\CN=4$ SYM on $\R^4$ has a corresponding gauge-invariant
state $|\CO \rangle = \lim_{r \rightarrow 0} \CO |0\rangle$ for
$\CN=4$ SYM on $\R \times S^3$ ($r$ being the radial coordinate of
$\R^4$), and vice versa, by the state/operator correspondence.}
\begin{equation}
\label{su2op}  \tr( A^{(1)}_1 A^{(1)}_2 \cdots A^{(1)}_{L_1} )\tr(
A^{(2)}_1 A^{(2)}_2 \cdots A^{(2)}_{L_2} ) \cdots \tr( A^{(k)}_1
A^{(k)}_2 \cdots A^{(k)}_{L_k} ) , \ A^{(i)}_{j} =Z,X
\end{equation}
This is in fact the so-called $SU(2)$ sector of recent interest in
the study of integrability of $\CN=4$ SYM
\cite{Minahan:2002ve,Beisert:2003tq,Beisert:2003jj,Beisert:2003yb,Beisert:2004hm}.
We can view this as a quantum mechanical subset of $\CN=4$ SYM in
the sense that all the states with covariant derivatives are
decoupled, which can be interpreted to mean that the modes
corresponding to moving around on the $S^3$ disappear, leaving us
with only one point.

Furthermore, as we show in \cite{Harmark:2006di}, the partition
function \eqref{genZ} reduces in the decoupling limit
\eqref{su2limit} to the partition function
\begin{equation}
\label{Zsu2} Z(\tilde{\beta}) = {\tr}_{\CH} \left(
e^{-\tilde{\beta} H} \right)
\end{equation}
with $H$ being the Hamiltonian
\begin{equation}
\label{Hsu2} H = D_0 + \tilde{\lambda} D_2
\end{equation}
Here $\tilde{\beta}=1/\tilde{T}$, thus we see that $SU(N)$ $\CN=4$
SYM on $\R \times S^3$ in the limit \eqref{su2limit} reduces to a
quantum mechanical theory with Hilbert space $\CH$ given by
\eqref{su2op} and with Hamiltonian \eqref{Hsu2}, with effective
temperature $\tilde{T}$. Moreover, $\tilde{\lambda}$ can be
regarded as the coupling of the theory, being a remnant of the 't
Hooft coupling of $\CN=4$ SYM. It is very interesting to observe
that we thus end up with a theory with two coupling constants:
$\tilde{\lambda}$ and $1/N$, both of which we can choose freely.
Indeed, since the $D_2$ term in \eqref{Hsu2} origins in the
one-loop correction to the scaling dimension, we have full
knowledge of the Hamiltonian \eqref{Hsu2} and we can in principle
compute $Z(\tilde{\beta})$ for any value of $\tilde{\lambda}$ and
$N$.

We can view the decoupling limit \eqref{su2limit} from the
alternative view point as a decoupling limit of non-thermal
$SU(N)$ $\CN=4$ SYM on $\R \times S^3$. Then the decoupling limit
is instead%
\footnote{When we write that $J_i$ is fixed we mean that all three
R-charges $J_1$, $J_2$ and $J_3$ are fixed.}
\begin{equation}
\label{altsu2limit} \epsilon \rightarrow 0 \spa
\frac{D-J}{\epsilon} \ \mbox{fixed} \spa \tilde{\lambda} \equiv
\frac{\lambda}{\epsilon} \ \mbox{fixed} \spa J_i \ \mbox{fixed}
\spa N \ \mbox{fixed}
\end{equation}
Note then that the effective Hamiltonian is $\lim_{\epsilon
\rightarrow 0} \frac{D-J}{\epsilon} = \tilde{\lambda} D_2$. We see
that this is in accordance with the Hamiltonian \eqref{Hsu2} since
we are restricting ourselves to be in a certain sector of fixed $J$.
We see that this limit is remarkably different from pp-wave limits
of $\CN=4$ SYM \cite{Berenstein:2002jq} in which one takes $J$ and
$N$ to go to infinity and instead fixes $D-J$. However, as we shall
see below we have an overlap between the two types of limits for a
particular pp-wave limit found in \cite{Bertolini:2002nr}.

\subsubsection*{The planar limit and the Heisenberg spin chain}

If we consider the planar limit $N \rightarrow \infty$ of $SU(N)$
$\CN=4$ SYM on $\R \times S^3$, we know from large $N$
factorization that the single-trace operators are decoupled from
the multi-trace operators. Therefore, in the planar limit, we can
regard single-trace operators of a certain length as states for a
spin chain where the letters correspond to the value of the spin
\cite{Minahan:2002ve}. In the $SU(2)$ sector, the single-trace
operators are linear combinations of
\begin{equation}
\label{singsu2op}  \tr( A_1 A_2 \cdots A_{L} ) , \ A_{i} =Z,X
\end{equation}
If we write $S_z = (J_1 - J_2)/2$ we see that each $Z$ has $S_z =
1/2$ and each $X$ has $S_z = -1/2$, thus we get an $SU(2)$ spin
chain. Furthermore, in the planar limit the $D_2$ term in
\eqref{Hsu2} is given by \cite{Minahan:2002ve,Beisert:2003tq}
\begin{equation}
D_2 = \frac{1}{2} \sum_{i=1}^L ( I_{i,i+1} - P_{i,i+1} )
\end{equation}
for a chain of length $L$, where $P_{i,i+1}$ is the permutation
operator acting on letters at position $i$ and $i+1$. From this
one can see that $\tilde{\lambda} D_2$ precisely is the
Hamiltonian for a ferromagnetic $XXX_{1/2}$ Heisenberg spin chain
of length $L$ \cite{Minahan:2002ve}. We can therefore write the
single-trace partition function as \cite{Harmark:2006di}
\begin{equation}
\label{ZST} Z_{\rm ST} (\tilde{\beta}) = \sum_{L=1}^\infty
e^{-\tilde{\beta} L} Z^{\rm (XXX)}_L ( \tilde{\beta} )
\end{equation}
where
\begin{equation}
\label{ZXXX} Z^{\rm (XXX)}_L ( \tilde{\beta} ) = {\tr}_L \left(
e^{-\tilde{\beta}\tilde{\lambda} D_2} \right)
\end{equation}
is the partition function for the ferromagnetic $XXX_{1/2}$
Heisenberg spin chain of length $L$ with Hamiltonian
$\tilde{\lambda} D_2$. Note that ${\tr}_L$ here refers to the
trace over single-trace operators with $J=L$ in the $SU(2)$
sector. The spin chain is required to be periodic and
translationally invariant in accordance with the cyclic symmetry
of single-trace operators. Using the standard relation between the
single-trace and multi-trace partition functions, we get that the
full partition function of planar $\CN=4$ SYM on $\R \times S^3$
in the limit \eqref{su2limit} is \cite{Harmark:2006di}
\begin{equation}
\label{planarZ} \log Z (\tilde{\beta}) = \sum_{n=1}^\infty
\sum_{L=1}^\infty \frac{1}{n} e^{- \tilde{\beta} nL} Z^{\rm
(XXX)}_L ( n \tilde{\beta} )
\end{equation}
Therefore, the partition function of planar $\CN=4$ SYM on $\R
\times S^3$ in the decoupling limit \eqref{su2limit} is given
exactly by \eqref{planarZ} from the partition function $Z^{\rm
(XXX)}_L(\tilde{\beta})$ of the ferromagnetic $XXX_{1/2}$
Heisenberg spin chain \cite{Harmark:2006di}.

\section{Gauge theory spectrum in decoupling limit}
\label{sec:spectrum}

In this section we find the large $\tilde{\lambda}$ and large $L$
limit of the spectrum of planar $\CN=4$ SYM on $\R \times S^3$ in
the decoupling limit \eqref{su2limit}.

From \eqref{Hsu2} we know that planar $\CN=4$ SYM on $\R \times
S^3$ in the limit \eqref{su2limit} has the Hamiltonian $L +
\tilde{\lambda} D_2$, for single-traces of length $L$. Therefore,
finding the spectrum of planar $\CN=4$ SYM in this decoupling
limit is identical to the problem of finding the spectrum of the
Heisenberg chain Hamiltonian $\tilde{\lambda} D_2$. The solution
to this for low energies is well-known. Nevertheless, we rederive
the spectrum in the following since we are interested in the case
where we have a degeneracy of the vacuum with respect to the total
spin. In our approach we employ a new way of putting in impurities
which seems more natural for this situation. It also makes a
direct construction of the eigenstates corresponding to the
spectrum possible.

We begin by noting that the large $\tilde{\lambda}$ limit of the
spectrum alternatively can be viewed as the low energy part of the
spectrum for finite $\tilde{\lambda}$, since the interacting term
in the Hamiltonian is $\tilde{\lambda} D_2$.

The low energy part of the spectrum of the ferromagnetic
Heisenberg chain consists of the ferromagnetic vacuum states plus
magnon excitations. The ferromagnetic vacua consist of all the
states for which $P_{i,i+1}$ has eigenvalue one for any
neighboring sites of the spin chain. One can make such a state for
each possible value of the total spin $S_z$, here given by
\begin{equation}
S_z = \frac{1}{2} ( J_1-J_2 )
\end{equation}
In detail we have that the vacuum state for a given length $L$ and
total spin $S_z$ is the totally symmetrized state
\cite{Harmark:2006di}
\begin{equation}
\label{vacua1} | S_z \rangle_L \sim \tr \big( \mbox{sym} ( Z^{J_1}
X^{J_2} ) \big)
\end{equation}
with $J_1 = \frac{1}{2}L + S_z$ and $J_2 = \frac{1}{2}L - S_z$. We
see thus that we have $L+1$ ferromagnetic vacua for a given length
$L$. As observed in \cite{Harmark:2006di}, the vacua
\eqref{vacua1} are precisely the chiral primary states with
$D_0=J$.

It will be useful below to have a more specific way of describing
the vacuum states. To this end, define $A_{1/2} = Z$ and $A_{-1/2}
=X$. Then we can write the basis of the $SU(2)$ sector as
\begin{equation}
\tr \big( A_{s(1)} \cdots A_{s(L)} \big)
\end{equation}
where $s(i) = \pm 1/2$ corresponds to having spin up or spin down.
Write
\begin{equation}
Q = \Big\{ s = \big(s(1),...,s(L) \big) \Big| \sum_{i=1}^L s(i) =
S_z \Big\}
\end{equation}
Then we have that the vacuum for a given value of $S_z$ and $L$ is
\begin{equation}
\label{vacua2} | S_z \rangle_L \sim  \sum_{s \in Q} \tr \big(
A_{s(1)} \cdots A_{s(L)} \big)
\end{equation}

Turning to the magnons, which are the low energy excitations of
the ferromagnetic vacua, we see that we cannot employ the usual
Bethe ansatz technique of putting $X$ impurities into a sea of
$Z$'s. This is due to the fact that we want to work in the limit
in which the number of excitations is much less than $L$, and
clearly it would take of order $L$ impurities to describe
excitations around vacua with $J_1 \ll J_2$. This difference to
the usual approach basically comes in because the $J_1=L$ vacuum
$\tr (Z^L)$ is not special, instead we have $L+1$ vacua which are
equally important.

Thus, we need a new way to put in impurities that does not change
the value of $S_z$. The way to do this becomes clearer if we think
of an impurity as the action of an operator on a particular site.
In particular changing a $Z$ at site number $l$ into an $X$ can be
thought of as the action of $S_-$ at site $l$. We instead want an
operator in the $SU(2)$ group that commutes with the total spin
$S_z$. Therefore, we propose that inserting an impurity
corresponds to the action of $S_z$ at a particular site $l$.%
\footnote{This way of constructing magnons is inspired from the
construction of gauge theory states in \cite{Bertolini:2002nr}.}

Consider the insertion of two impurities. Define $S_{z,l}$ as the
action of $\frac{1}{2}(Z \partial_Z - X \partial_X )$ on the site
number $l$. We can then write the insertion of two impurities at
sites $l_1$ and $l_2$ in the vacuum state $|S_z \rangle_L$ as
\begin{equation}
| l_1, l_2 ; S_z \rangle_L = S_{z,l_1}  S_{z,l_2} | S_z \rangle_L
\end{equation}
Using the form \eqref{vacua2} for the vacuum states, we see that
this corresponds to
\begin{equation}
| l_1, l_2 ; S_z \rangle_L \sim \sum_{s \in Q} s(l_1) s(l_2) \tr
\big( A_{s(1)} \cdots A_{s(L)} \big)
\end{equation}
We now want to find an eigenstate of the Hamiltonian
$\tilde{\lambda} D_2$ with two impurities. Write
\begin{equation}
| \Psi \rangle = \sum_{1\leq l_1 \leq l_2 \leq L} \Psi(l_1,l_2) |
l_1, l_2 ; S_z \rangle_L
\end{equation}
The task is then to find $\Psi(l_1,l_2)$ such that
\begin{equation}
\label{eigeq} \tilde{\lambda} D_2 | \Psi \rangle = \tilde{\lambda}
\CE | \Psi \rangle
\end{equation}
To this end, we employ the Bethe ansatz
\begin{equation}
\Psi (l_1,l_2) = e^{ip_1 l_1 + i p_2 l_2 } A_{12} + e^{ip_2 l_1 +
i p_1 l_2 } A_{21}
\end{equation}
It is not hard to see that the eigenvalue equation \eqref{eigeq}
then gives
\begin{equation}
\CE = 2 \sum_{k=1}^2 \sin^2 \Big( \frac{p_k}{2} \Big) \spa
S(p_1,p_2) \equiv \frac{A_{12}}{A_{21}} = - \frac{1+
e^{i(p_1+p_2)} - 2e^{ip_1}}{1+e^{i(p_1+p_2)} - 2e^{ip_2}}
\end{equation}
Periodicity of the spin chain instead requires
\begin{equation}
e^{ip_1 L} = S(p_1,p_2) \spa e^{ip_2 L} = S(p_2,p_1)
\end{equation}
Furthermore, the cyclicity of the trace requires $p_1 + p_2 = 0$.
Using these conditions, one can easily determine the spectrum for
two impurities.

Considering the general case of inserting $q$ impurities, we can
use the integrability of the Heisenberg chain to find the
spectrum, giving
\begin{equation}
\CE = 2 \sum_{i=1}^q \sin^2 \Big( \frac{p_i}{2} \Big)
\end{equation}
\begin{equation}
\label{betheeq} e^{ip_k L} = \prod_{j=1,j\neq k}^q S(p_k,p_j )
\spa S(p_k,p_j ) = - \frac{1+ e^{i(p_k+p_j)} -
2e^{ip_k}}{1+e^{i(p_k+p_j)} - 2e^{ip_j}}
\end{equation}
\begin{equation}
\label{totmom} \sum_{i=1}^q p_i = 0
\end{equation}
where \eqref{totmom} is due to the cyclicity of the trace. Taking
the logarithm of \eqref{betheeq} we have
\begin{equation}
\label{betheeq2} p_k - \frac{2\pi n_k}{L} =  - \frac{i}{L}
\sum_{j=1,j\neq k}^q \log S(p_k,p_j )
\end{equation}
where $n_k$ is an integer. The leading order solution for large
$L$ is
\begin{equation}
p_k = \frac{2\pi n_k}{L} + \CO ( L^{-2} )
\end{equation}
giving the spectrum
\begin{equation}
\CE = \frac{2 \pi^2}{L^2} \sum_{i=1}^q n_i^2 \spa \sum_{i=1}^q n_i
=0
\end{equation}
Denoting the number of $n_i$ which are equal to a particular
integer $k$ as $M_k$, we can write this spectrum as
\begin{equation}
\label{spectrum} \CE = \frac{2 \pi^2}{L^2} \sum_{n \neq 0} n^2 M_n
\spa \sum_{n \neq 0} n M_n = 0
\end{equation}
This is the low energy spectrum of the Heisenberg spin chain with
spin chain Hamiltonian $D_2$. Using now that for a single-trace
operator of length $L$, the eigenvalue of the Hamiltonian
$H=D_0+\tilde{\lambda} D_2$ is $L + \tilde{\lambda} \CE$, we see
that the Hamiltonian $H$ has the spectrum
\begin{equation}
\label{Hspectrum} H-L = \frac{2 \pi^2 \tilde{\lambda}}{L^2}
\sum_{n \neq 0} n^2 M_n \spa \sum_{n \neq 0} n M_n = 0
\end{equation}
This is the large $\tilde{\lambda}$ and large $L$ limit of the
spectrum of single-trace operators in planar $\CN=4$ SYM on $\R
\times S^3$ in the decoupling limit \eqref{su2limit}.

We see that the spectrum \eqref{Hspectrum} is string-like, even
though we are in weakly coupled gauge theory. This is in contrast
with previous approaches to find a string-like spectrum in $\CN=4$
SYM on $\R \times S^3$, since those approaches rely on having
$\lambda$ large in order to decouple gauge theory states which are
not near the chiral primary states. Thus, in this sense, the
spectrum \eqref{spectrum} is the first example of a string-like
spectrum found in weakly-coupled $\CN=4$ SYM. As we shall see in
Section \ref{sec:conpp}, the resemblance to a string-spectrum is
not accidental, and we can in fact map it to a spectrum of string
states in a decoupling limit of strings on a pp-wave.

\section{Gauge theory Hagedorn temperature from the Heisenberg chain}
\label{sec:gaugehag}

In this section we consider the Hagedorn temperature of planar
$\CN=4$ SYM on $\R \times S^3$ in the decoupling limit
\eqref{su2limit} from a general perspective, and we find a
relation between the Hagedorn temperature as function of
$\tilde{\lambda}$ and the thermodynamics of the Heisenberg chain
in the thermodynamic limit. We use this general connection to find
the Hagedorn temperature for small and large $\tilde{\lambda}$.

\subsection{General considerations}

From \eqref{planarZ} and \eqref{ZXXX} we have that the full
partition function of planar $\CN=4$ SYM on $\R \times S^3$ in the
decoupling limit \eqref{su2limit} is
\begin{equation}
\label{thelogZ} \log Z (\tilde{\beta}) = \sum_{n=1}^\infty
\sum_{L=1}^\infty \frac{1}{n} e^{- \tilde{\beta} nL} {\tr}_L
\left( e^{-n\tilde{\beta}\tilde{\lambda} D_2} \right)
\end{equation}
Define now the function $V(t)$ by
\begin{equation}
\label{defV} V(t) \equiv \lim_{L \rightarrow \infty} \frac{1}{L}
\log \left[ {\tr}_L \left( e^{- t^{-1} D_2} \right) \right]
\end{equation}
This limit is well-defined since the thermodynamic limit of the free
energy per site $f(t)$ at temperature $t$ for the Heisenberg chain
is related to $V(t)$ by
\begin{equation}
f(t) = - t V(t)
\end{equation}
Note that here the Hamiltonian of the ferromagnetic Heisenberg
chain is $D_2$. We notice now that for large $L$
\begin{equation}
e^{- \tilde{\beta} nL} {\tr}_L \left(
e^{-n\tilde{\beta}\tilde{\lambda} D_2} \right) \simeq \exp \left(
- n L \tilde{\beta} + L V \big[ (n \tilde{\beta}
\tilde{\lambda})^{-1} \big] \right)
\end{equation}
Therefore, for $n=1$ we see that we reach a singularity if
$\tilde{\beta}$ decreases to $\tilde{\beta}_H$ given by%
\footnote{Note that there is a singularity for each value of $n$,
but the $n=1$ singularity is the first one that is reached as one
decreases $\tilde{\beta}$ from infinity. This is seen using that
$V(t)$ is a monotonically increasing function of $t$.}
\begin{equation}
\label{genhag} \tilde{\beta}_H = V \left( (\tilde{\beta}_H
\tilde{\lambda})^{-1} \right)
\end{equation}
This is the Hagedorn temperature for general $\tilde{\lambda}$.
Thus, we have obtained a direct connection between the
thermodynamics of the Heisenberg chain in the thermodynamic limit
and the Hagedorn temperature.

We see now immediately from Eq.~\eqref{genhag} that the Hagedorn
temperature for $\tilde{\lambda} \ll 1$ is obtained from the high
temperature limit $t \gg 1$ of the Heisenberg chain, while for
$\tilde{\lambda} \gg 1$ the Hagedorn temperature is obtained from
the low temperature limit $t \ll 1$. In the following we use this
to obtain the Hagedorn temperature in these two regimes.

\subsection{Hagedorn temperature for small $\tilde{\lambda}$}

If we consider $t \rightarrow \infty$ in \eqref{defV} we see that
we can find the Hagedorn temperature from ${\tr}_L (1)$. This
corresponds to counting the number of independent single-trace
operators of length $L$. This is less than $2^L$ but also bigger
than $2^L/L$ since the cyclic symmetry of the trace can at most
relate $L$ states to each other. For large $L$ we have therefore
to leading order ${\tr}_L (1) \simeq 2^L$. Inserting that in
\eqref{defV} we see that $V(t) \rightarrow \log 2$ for $t
\rightarrow \infty$. This corresponds to $\tilde{\beta}_H = \log
2$ which is the correct Hagedorn temperature for the free $SU(2)$
sector.

We can also find the first correction to the Hagedorn temperature
for small $\tilde{\lambda}$ in this fashion. For large $t$ we see
that
\begin{equation}
V(t) = \lim_{L \rightarrow \infty} \frac{1}{L} \left[ \log {\tr}_L
(1) - t^{-1} \frac{{\tr}_L (D_2)}{{\tr}_L (1)} \right]
\end{equation}
It is not hard to see that for large $L$
\begin{equation}
\frac{{\tr}_L (D_2)}{{\tr}_L (1)} \simeq \frac{L}{4}
\end{equation}
Therefore, we get
\begin{equation}
\label{Vhight1} V(t) = \log 2 - \frac{1}{4t} + \CO (t^{-2} )
\end{equation}
for large $t$. We see from \eqref{Vhight1} that $-t V(t)$ indeed
is the previously computed high temperature limit of the free
energy per site for the Heisenberg chain \cite{takahashi}.
Inserting \eqref{Vhight1} into \eqref{genhag} we get
\begin{equation}
\tilde{T}_H =\frac{1}{\log{2}} +\frac{1}{4 \log 2}\tilde{\lambda}
+ \CO(\tilde{\lambda}^2)
\end{equation}
which precisely matches the Hagedorn temperature found previously in
\cite{Spradlin:2004pp,Harmark:2006di}. Note that the above
computation of the Hagedorn temperature completely circumvents the
somewhat complicated computation of the full single-trace partition
function.

A much more powerful method of obtaining the high temperature
behavior of the Heisenberg chain has been found in
\cite{PhysRevLett.89.117201}. The result is that $V(t)$ as defined
in \eqref{defV} can be found from the integral equation
\begin{equation}
\label{ueq} u(x) = 2 + \oint_C \frac{dy}{2\pi i} \left\{
\frac{1}{x-y-2i} \exp \left[ - \frac{2t^{-1}}{y(y+2i)} \right] +
\frac{1}{x-y+2i} \exp \left[ - \frac{2t^{-1}}{y(y-2i)} \right]
\right\} \frac{1}{u(y)}
\end{equation}
where $C$ is a loop around the origin directed counterclockwise.
$V(t)$ is then determined as
\begin{equation}
V(t) = \log \big[ u (0) \big]
\end{equation}
One can then make a systematic high energy expansion of $u(x)$ in
powers of $t^{-1}$ as
\begin{equation}
\log \big[ u(x) \big] = \sum_{k=0}^\infty u_k(x) t^{-k}
\end{equation}
Using \eqref{ueq} we can now determine $u(x)$ order by order in
$t^{-1}$. This gives the high temperature expansion of $V(t)$ to
order $t^{-5}$
\begin{equation}
\label{Vhight2} V(t) = \log 2 - \frac{1}{4t} + \frac{3}{32t^2} -
\frac{1}{64 t^3} - \frac{5}{1024 t^4} + \frac{3}{1024 t^5} + \CO
(t^{-6})
\end{equation}
for large $t$. Inserting \eqref{Vhight2} into \eqref{genhag} we
get%
\footnote{Note that the $\tilde{\lambda}^2$ term matches the
$D_2^2$ contribution to the $\lambda^2$ correction for the
Hagedorn temperature in the $SU(2)$ sector found in
\cite{Gomez-Reino:2005bq}.}
\begin{equation}
\label{THsmall} \begin{array}{rcl} \ds \tilde{T}_H &=& \ds
\frac{1}{\log 2} + \frac{1}{4\log 2} \tilde{\lambda} -
\frac{3}{32} \tilde{\lambda}^2 + \left( \frac{3}{128} + \frac{\log
2}{64} \right) \tilde{\lambda}^3  + \left( - \frac{3}{512} -
\frac{17\log 2}{1024} + \frac{5 (\log 2)^2}{1024} \right)
\tilde{\lambda}^4 \\[4mm] && \ds + \left( \frac{3}{2048} + \frac{39 \log 2}{4096}
+ \frac{3(\log 2)^2}{4096} - \frac{3(\log 2)^3}{1024} \right)
\tilde{\lambda}^5 + \CO (\tilde{\lambda}^6)
\end{array}
\end{equation}
for small $\tilde{\lambda}$. It is straightforward to extend this
to higher orders in $\tilde{\lambda}$, e.g. from the results of
\cite{PhysRevLett.89.117201} one can find $V(t)$ to order
$t^{-50}$ and thereby $\tilde{T}_H$ to order
$\tilde{\lambda}^{50}$.

\subsection{Hagedorn temperature for large $\tilde{\lambda}$}
\label{sec:hagtempll}

As stated above, we see from \eqref{genhag} that the Hagedorn
temperature for large $\tilde{\lambda}$ is given from low
temperature limit of the ferromagnetic Heisenberg chain.
Therefore, to compute the Hagedorn temperature in this limit, we
should use the low energy spectrum \eqref{spectrum} of the
Heisenberg chain to compute $V(t)$ for small $t$. Inserting the
spectrum \eqref{spectrum} in the partition function for the
Heisenberg chain, we see that for large $L$ and small $t$ we have
\begin{equation}
{\tr}_L \left( e^{- t^{-1} D_2 } \right) = L
\sum_{\{M_n\}}\int_{-1/2}^{1/2}du \exp \left(- \frac{2\pi^2}{t
L^2} \sum_{n\neq 0} n^2 M_n+ 2\pi i u \sum_{n\neq 0} n M_n \right)
\end{equation}
where the integration over $u$ is introduced to impose the
cyclicity constraint in the spectrum \eqref{spectrum}. The $L$
factor is due to the $L+1$ different vacua for a given $L$.
Evaluating the sums over the $M_n$'s (the sum range being from
zero to infinity) we get
\begin{equation}
\label{Zsmallt} \begin{array}{rcl} \ds {\tr}_L \left( e^{- t^{-1}
D_2 } \right) &=& \ds L \int_{-1/2}^{1/2} du \prod_{n \neq 0}
\left[ 1- \exp \left( - \frac{2\pi^2}{tL^2} n^2 + 2\pi i u n
\right) \right]^{-1}
\\[3mm] \ds &=& \ds L \int_{-1/2}^{1/2} du \left| G \left(
\frac{2\pi}{tL^2} , 2\pi u \right) \right|^2
\end{array}
\end{equation}
where $G(a,b)$ is the generating function defined by
Eq.~\eqref{genfunc} in Appendix \ref{app:Asy}. We want to extract
from \eqref{Zsmallt} the part that diverges for $L \rightarrow
\infty$. Using the analysis of Appendix \ref{app:Asy} we get that
the leading contribution to this divergence is from $u=0$, which
using Eq.~\eqref{Ga0} is seen to give
\begin{equation}
\label{divcon} {\tr}_L \left( e^{- t^{-1} D_2 } \right) \sim \exp
\left\{ L\, \zeta \Big( \frac{3}{2} \Big) \sqrt{\frac{t}{2\pi}}\,
\right\}
\end{equation}
for $L \rightarrow \infty$. Here $\zeta(x)$ is the Riemann zeta
function. Inserting \eqref{divcon} into \eqref{Zsmallt} and
\eqref{defV} we get
\begin{equation}
\label{Vlowt} V(t) = \zeth \sqrt{\frac{t}{2\pi}}
\end{equation}
for $t \ll 1$. This result is the same as the analytically obtained
result \cite{PhysRevLett.58.168,takahashi} for the low energy limit
of the free energy $-tV(t)$ for the Heisenberg chain. As we discuss
further below, it is also consistent with numerical calculations
\cite{PhysRevLett.54.2131,PhysRevB.33.4880,JPhysSocJpn.54.2808,JPhysSocJpn.55.2024}.

Applying now the result \eqref{Vlowt} to Eq.~\eqref{genhag}, we
get the Hagedorn temperature
\begin{equation}
\tilde{T}_H = (2\pi)^{\frac{1}{3}} \left[ \zeth
\right]^{-\frac{2}{3} } \tilde{\lambda}^{\frac{1}{3}}
\label{haggauge}
\end{equation}
for $\tilde{\lambda} \gg 1$. This is the Hagedorn temperature of
planar $\CN=4$ SYM on $\R \times S^3$ in the decoupling limit
\eqref{su2limit} for large $\tilde{\lambda}$. We see that the
Hagedorn temperature \eqref{haggauge} goes to infinity for
$\tilde{\lambda} \rightarrow \infty$. This is consistent with the
fact that for $\tilde{\lambda} \rightarrow \infty$ all other states
except the chiral primary states decouple, and the partition
function ends up being a sum only over the chiral primaries, which
means that we should not expect the presence of a Hagedorn
singularity in this limit.

As stated above, the result \eqref{Vlowt} obtained for the low
temperature limit of $V(t)$ is the same as that obtained for the
ferromagnetic Heisenberg chain in
\cite{PhysRevLett.58.168,takahashi}, where also the next order of
$V(t)$ has been computed
\begin{equation}
\label{Vlowt2} V(t) = \zeth \sqrt{\frac{t}{2\pi}} - t +
\CO(t^{3/2} )
\end{equation}
for $t \ll 1$. This result is consistent with numerical
calculations, which reveals
\cite{PhysRevLett.54.2131,PhysRevB.33.4880,JPhysSocJpn.54.2808,JPhysSocJpn.55.2024}
\begin{equation}
\label{Vlowt3} V(t) = 1.042 \sqrt{t} - 1.00\, t + \CO(t^{3/2})
\end{equation}
for $t \ll 1$. Using now \eqref{Vlowt2} in \eqref{genhag} we find
the following correction to the Hagedorn temperature
\begin{equation}
\label{haggauge2} \tilde{T}_H = \frac{ (2\pi)^{1/3} }{ \zeta \big(
\frac{3}{2} \big)^{2/3 } } \tilde{\lambda}^{1/3} + \frac{4\pi}{3
\zeta\big( \frac{3}{2} \big)^2} + \CO( \tilde{\lambda}^{-1/3} )
\end{equation}
for large $\tilde{\lambda}$.

\section{Decoupling limit of string theory on $\ads_5 \times S^5$}
\label{sec:adslim}

As reviewed in Section \ref{sec:declim}, thermal $\CN=4$ SYM on
$\R \times S^3$ decouples to $SU(2)$ sector in the decoupling
limit \eqref{su2limit} \cite{Harmark:2006di}. We consider in this
section the corresponding limit that one obtains for type IIB
string theory on $\ads_5 \times S^5$ by employing the AdS/CFT
duality \cite{Maldacena:1997re,Gubser:1998bc,Witten:1998qj}.

We consider type IIB string theory on the $\ads_5 \times S^5$
background given by the metric
\begin{equation}
\label{adsmet} ds^2 = R^2 \left[ - \cosh^2 \rho dt^2 + d\rho^2 +
\sinh^2 \rho d{\Omega'_3}^2 + d\theta^2 + \sin^2 \theta d\alpha^2
+ \cos^2 \theta d\Omega_3^2 \right]
\end{equation}
and the five-form Ramond-Ramond field strength
\begin{equation}
\label{adsF5} F_{(5)} = 2 R^4 ( \cosh \rho \sinh^3 \rho dt d\rho
d\Omega_3' +
 \sin \theta \cos^3 \theta d\theta d\alpha d\Omega_3 )
\end{equation}
The AdS/CFT correspondence then fixes that $R^4 = 4\pi g_s l_s^4
N$ and $\gym^2 = 4\pi g_s$, where $g_s$ is the string coupling and
$l_s$ is the string length. $\gym^2$ and $N$ are the gauge
coupling and rank of $SU(N)$ as defined in Section
\ref{sec:declim}. With this, we see that we have the following
dictionary between the gauge theory quantities $\lambda$ and $N$,
and the string theory quantities $g_s$, $l_s$ and the AdS radius
$R$
\begin{equation}
\label{adsdic} T_{\rm str} \equiv \frac{R^2}{4\pi l_s^2} =
\frac{1}{2} \sqrt{\lambda} \spa g_s = \frac{\pi \lambda}{N}
\end{equation}
where $T_{\rm str}$ is the string tension for a fundamental string
in the $\ads_5 \times S^5$ background
\eqref{adsmet}-\eqref{adsF5}.

\subsubsection*{Decoupling limit for strings on $\ads_5 \times
S^5$ and induced gauge/string duality}

We can now translate the decoupling limit reviewed in Section
\ref{sec:declim}. We consider first the non-thermal version of the
decoupling limit given by \eqref{altsu2limit}. This limit
translates into the following limit of type IIB string theory on
the $\ads_5 \times S^5$ background \eqref{adsmet}-\eqref{adsF5}
\begin{equation}
\label{adslimit} \epsilon \rightarrow 0 \spa \tilde{H} \equiv
\frac{E-J}{\epsilon} \ \mbox{fixed} \spa \tilde{T}_{\rm str}
\equiv \frac{T_{\rm str}}{\sqrt{\epsilon}} \ \mbox{fixed} \spa
\tilde{g}_s \equiv \frac{g_s}{\epsilon}\ \mbox{fixed} \spa J_i\
\mbox{fixed}
\end{equation}
Here $E$ is the energy of the string while $J_i$, $i=1,2,3$, are
the three angular momenta for the five-sphere corresponding to the
three R-charges of $\CN=4$ SYM. The energy $E$ for a string state
is equal to the scaling dimension $D$ of a gauge theory state of
$\CN=4$ SYM on $\R \times S^3$ since we set the radius of the
three-sphere to one. Note furthermore that we have defined $J =
J_1+J_2$.

We see that in this limit we scale the energies in such a way that
in free string theory ($g_s=0$) only string states for which $E-J
\sim T_{\rm str}^2$ as $T_{\rm str} \rightarrow 0$ can survive. As
in the gauge theory, we can regard this as a limit in which we
look at small excitations near the BPS states with $E=J$. Note
that even for $g_s=0$ the obtained tree-level string theory is
non-trivial since we have an effective string tension
$\tilde{T}_{\rm str}$.

It is interesting to observe that in the limit \eqref{adslimit} the
string coupling goes to zero. From this and the corresponding gauge
theory limit \eqref{altsu2limit}, we see that the AdS/CFT
correspondence in this limit necessarily becomes a duality between
weakly coupled $\CN=4$ SYM and weakly coupled string theory.

After taking the limit \eqref{altsu2limit} of $\CN=4$ SYM and the
limit \eqref{adslimit} of string theory on $\ads_5 \times S^5$,
the AdS/CFT duality induces a duality between the decoupled
sectors on the gauge theory and string theory sides. From the two
limits \eqref{altsu2limit} and \eqref{adslimit} we see that we
obtain a dictionary for this induced duality relating the
quantities we keep finite in the limits:
\begin{equation}
\label{inddic} \tilde{T}_{\rm str} = \frac{1}{2}
\sqrt{\tilde{\lambda}} \spa \tilde{g}_s = \frac{\pi
\tilde{\lambda}}{N}
\end{equation}
We see that this induced dictionary perfectly mirrors the original
AdS/CFT dictionary \eqref{adsdic}.

Finally, we note also that the string tension $T_{\rm str}$ goes
to zero. Zero tension limits of string theory on $\ads_5 \times
S^5$ have previously been connected to higher-spin theories.
However, here we know from the gauge theory side that only a
particular sector of the theory survives the limit.

\subsubsection*{Decoupling limit of thermal partition function
for strings on $\ads_5\times S^5$}

If we consider instead a gas of strings in the $\ads_5 \times S^5$
background \eqref{adsmet}-\eqref{adsF5} we can write the general
partition function as
\begin{equation}
Z(\beta,\Omega_i) = \tr \left( e^{-\beta E + \beta \sum_{i=1}^3
\Omega_i J_i} \right)
\end{equation}
where $J_i$, $i=1,2,3$, are the angular momenta and $\Omega_i$,
$i=1,2,3$, are the corresponding angular velocities. Here we trace
over all the multi-string states. Just like on the gauge theory
side we consider here the only special case
$(\Omega_1,\Omega_2,\Omega_3)=(\Omega,\Omega,0)$. Therefore, the
partition function can be written
\begin{equation}
\label{partfct} Z(\beta,\Omega_i) = \tr \left( e^{-\beta E + \beta
\Omega J} \right)
\end{equation}
where $J = J_1+J_2$. We now want to consider the region close to
the critical point $(T,\Omega)=(0,1)$. We notice first that we can
rewrite the weight factor in \eqref{partfct} as
\begin{equation}
\label{wfac} e^{-\beta E + \beta \Omega J} = e^{ - \beta
(1-\Omega) J - \beta (1-\Omega) \frac{E-J}{1-\Omega} }
\end{equation}
From the gauge theory decoupling limit \eqref{su2limit} and the
string theory decoupling limit \eqref{adslimit} it is then clear
that the appropriate limit for a string gas is
\begin{equation}
\label{adsthermlim} \begin{array}{c} \ds T \rightarrow 0 \spa
\Omega \rightarrow 1 \spa \tilde{T} = \frac{T}{1-\Omega} \
\mbox{fixed} \spa \tilde{H} \equiv \frac{E-J}{1-\Omega} \
\mbox{fixed} \\[3mm] \ds \tilde{T}_{\rm str} \equiv \frac{T_{\rm
str}}{\sqrt{1-\Omega}} \ \mbox{fixed} \spa \tilde{g}_s \equiv
\frac{g_s}{1-\Omega}\ \mbox{fixed} \spa J_i\ \mbox{fixed}
\end{array}
\end{equation}
Using \eqref{partfct} and \eqref{wfac} the partition function for
the string gas becomes
\begin{equation}
Z (\tilde{\beta} ) = {\tr}_{\CM_s} \left( e^{-\tilde{\beta} ( J +
\tilde{H}) } \right)
\end{equation}
where $\CM_s$ is defined as the set of all multi-string states
that survive the limit \eqref{adsthermlim}. We see that in the
limit \eqref{adsthermlim} we effectively end up with a theory for
a string gas of temperature $\tilde{T}$ and energies given by
$J+\tilde{H}$, and with a reduced set of string states compared to
the full string theory on $\ads_5\times S^5$.

\section{Connection to pp-wave with flat direction}
\label{sec:conpp}

In Section \ref{sec:adslim} we found a decoupling limit of string
theory on $\ads_5 \times S^5$ which is dual to the $SU(2)$
decoupling limit of $\CN=4$ SYM reviewed in Section
\ref{sec:declim}. We do not know a first quantization of string
theory on $\ads_5 \times S^5$. Therefore, we consider instead
taking the decoupling limit \eqref{adslimit} for string theory on
a particular pp-wave background, obtained from $\ads_5 \times S^5$
by a Penrose limit. As we explain in the following, this pp-wave
background is particularly well-suited for this limit, and we find
indeed a successful match of the string theory and gauge theory
spectra.

\subsection{Penrose limit for pp-wave with flat direction}

We begin this section by employing a Penrose limit of $\ads_5
\times S^5$ found in \cite{Bertolini:2002nr} giving rise to a
maximally supersymmetric pp-wave background with a flat direction.
 It is
important to note that the Penrose limit is implemented in a
slightly different manner here than in \cite{Bertolini:2002nr} in
order to be consistent with the decoupling limit \eqref{adslimit}
for strings on $\ads_5\times S^5$. We explain in Section
\ref{sec:declimpp} why the Penrose limit of
\cite{Bertolini:2002nr} has the right features for the decoupling
limit \eqref{adslimit} that we are going to implement.

We begin by considering the $\ads_5 \times S^5$ background
\eqref{adsmet}-\eqref{adsF5}. We see from the decoupling limit
\eqref{adslimit} that the AdS radius $R$ goes to zero like
$\epsilon^{1/4}$ in the limit. We define therefore a rescaled AdS
radius $\tilde{R}$ as follows
\begin{equation}
\label{Rtilde} \tilde{R}^4 = \frac{R^4}{\epsilon}
\end{equation}
Consider now the three-sphere $\Omega_3$ part of the metric
\eqref{adsmet}. Following \cite{Bertolini:2002nr}, we can
parameterize the three-sphere embedded in the five-sphere as
\begin{equation}
\label{3sph} d\Omega_3^2 = d\psi^2 + \sin^2 \psi d\phi^2 + \cos^2
\psi d\chi^2 = d\psi^2 + d\phi_-^2 + d\phi_+^2 + 2 \cos (2\psi)
d\phi_- d\phi_+
\end{equation}
where we defined the angles $\phi_\pm$ as
\begin{equation}
\phi_\pm = \frac{\chi \pm \phi}{2}
\end{equation}
Define now the coordinates $x^+$, $x^-$, $x^1$, $x^2$, $r$,
$\tilde{r}$ by
\begin{equation}
x^- = \frac{1}{2} \mu \tilde{R}^2 (t-\phi_+) \spa x^+ =
\frac{1}{2\mu} (t+\phi_+)
\end{equation}
\begin{equation}
x^1 = \tilde{R} \phi_- \spa x^2 = \tilde{R} \left( \psi -
\frac{\pi}{4} \right) \spa r = \tilde{R} \rho \spa \tilde{r} =
\tilde{R} \theta
\end{equation}
Note that these coordinates are defined in terms of the rescaled
AdS radius $\tilde{R}$. We then take the Penrose limit of the
$\ads_5 \times S^5$ background \eqref{adsmet}-\eqref{adsF5} given
by \cite{Bertolini:2002nr}
\begin{equation}
\label{Plimc} \tilde{R} \rightarrow \infty \spa x^+, \ x^-, \ x^1
, \ x^2 , \ r , \ \tilde{r} , \ \alpha \ \mbox{fixed}
\end{equation}
This gives the following pp-wave background with 32
supersymmetries
\begin{equation}
\label{ppmet} \frac{ds^2}{\sqrt{\epsilon}} = - 4dx^+ dx^- - \mu^2
\sum_{I=3}^8 x^I x^I (dx^+)^2 + \sum_{i=1}^8 dx^i dx^i - 4 \mu x^2
dx^1 dx^+
\end{equation}
\begin{equation}
\label{ppF5} \frac{F_{(5)}}{\epsilon} = 2\mu dx^+ (dx^1 dx^2 dx^3
dx^4 + dx^5 dx^6 dx^7 dx^8 )
\end{equation}
This background was first found in \cite{Michelson:2002wa}.%
\footnote{The pp-wave background \eqref{ppmet}-\eqref{ppF5} is
related to the maximally supersymmetric pp-wave background of
\cite{Blau:2001ne,Berenstein:2002jq} by a coordinate
transformation \cite{Michelson:2002wa,Bertolini:2002nr}. Even so,
as we shall see in the following, the physics of this pp-wave is
rather different, which basically origins in the fact that the
coordinate transformation between them depends on $x^+$, i.e. it
is time-dependent. See \cite{Bertolini:2002nr} for more comments
on this.} Here $x^3,x^4$ are defined by $x^3+ix^4 = \tilde{r}
e^{i\alpha}$ and $x^5,...,x^8$ are defined by $r^2 = \sum_{I=5}^8
(x^I)^2$ and $dr^2 + r^2 d{\Omega'_3}^2 =\sum_{I=5}^8 (dx^I)^2$.
We see that the fact that we employed the rescaled AdS radius in
the Penrose limit give rise to factors of $\epsilon$ in the metric
and five-form field strength. This will be important below.

It is important to note that the pp-wave background
\eqref{ppmet}-\eqref{ppF5} has the special feature that $x^1$ is
an explicit isometry of the pp-wave
\cite{Michelson:2002wa,Bertolini:2002nr}, hence we call this
background a pp-wave with a flat direction.

In terms of the generators, we see that in the Penrose limit
\eqref{Plimc} we have
\begin{equation}
\label{gene} H_{\rm lc} = \sqrt{\epsilon}\, \mu ( E- J ) \spa p^+
= \frac{E+J}{2\mu R^2} \spa p_1 = \frac{2S_z}{\tilde{R}}
\end{equation}
where $H_{\rm lc}$ is the light-cone Hamiltonian, $p^+$ is the
light-cone momentum and $p_1$ is the momentum along the $x^1$
direction. Here $J_1 = \frac{1}{2} J + S_z$ and $J_2 = \frac{1}{2}
J - S_z$ are the angular momenta of the strings on the
three-sphere \eqref{3sph}.

From \cite{Michelson:2002wa,Bertolini:2002nr} we have that the
strings can be quantized in the light-cone gauge with the
following spectrum of the light-cone Hamiltonian $H_{\rm lc}$
\begin{equation}
\label{fullstrspec}
\begin{array}{rcl} \ds \frac{l_s^2 p^+}{\sqrt{\epsilon}} H_{\rm lc} &=&\ds 2 f N_0 + \sum_{n\neq 0} \left[
(\omega_n+f) N_n + (\omega_n - f) M_n \right] + \sum_{n\in \Z}
\sum_{I=3}^8 \omega_n N_n^{(I)} \\[5mm] && \ds + \sum_{n\in \Z}
\left[ \sum_{b=1}^4 \left( \omega_n - \frac{1}{2} f \right)
F_n^{(b)} +\sum_{b=5}^8 \left( \omega_n + \frac{1}{2} f \right)
F_n^{(b)} \right]
\end{array}
\end{equation}
with level matching condition
\begin{equation}
\label{levmat} \sum_{n\neq0} n \left[ N_n + M_n + \sum_{I=3}^8
N^{(I)}_n + \sum_{b=1}^8 F^{(b)}_n \right] = 0
\end{equation}
and where we have defined
\begin{equation}
f = \mu l_s^2 p^+ \spa \omega_n = \sqrt{n^2 + f^2}
\end{equation}
Here $N_n^{(I)}$, $I=3,...,8$ and $n\in Z$, are the number
operators for bosonic excitations for the six directions
$x^3,...,x^8$, while $N_n$, $n \in \Z$, and $M_n$, $n \neq 0$, are
the number operators for the two directions $x^1$ and $x^2$.
$F_n^{(b)}$, $b=1,...,8$ and $n\in \Z$, are the number operators
for the fermions. Note that the presence of the flat direction
$x^1$ of the pp-wave is responsible for the fact that we only have
seven bosonic zero modes $N_0$ and $N^{(3)}_0,...,N^{(8)}_0$.

It is important to note that the vacua for the string spectrum are
degenerate with respect to the eigenvalues of the momentum $p_1$
along the flat direction. I.e. we have a vacuum $|0,p_1,p^+ \rangle$
for each value of $p_1$, and given any particular vacuum $|0,p_1,p^+
\rangle$ we have the spectrum \eqref{fullstrspec} of string
excitations.

\subsection{Decoupling limit of pp-wave spectrum and matching of
spectra} \label{sec:declimpp}

We can now explain why the pp-wave background
\eqref{ppmet}-\eqref{ppF5} is relevant for our decoupling limit
\eqref{adslimit} for strings on $\ads_5 \times S^5$. We see from
\eqref{gene} that the Penrose limit \eqref{Plimc} corresponds to a
limit in which $J=J_1+J_2 \rightarrow \infty$ while $E-J$ is
fixed. Thus, we keep all excitations that have a finite value of
$E-J$. In particular, we keep any excitation which has a small
$E-J$ and which is still present for large $J$.

Another argument why the pp-wave background
\eqref{ppmet}-\eqref{ppF5} is suitable for our considerations is
that the light-cone vacua $H_{\rm lc} =0$ correspond to $1/2$ BPS
states with $E=J$. These $1/2$ BPS states are mapped to the chiral
primary states of $\CN=4$ SYM with $D=J$, which precisely
correspond to the vacua on the gauge theory side.

We now implement the decoupling limit \eqref{adslimit} on the
pp-wave background \eqref{ppmet}-\eqref{ppF5}. Notice first that
we want to keep $p^+$ fixed in the decoupling limit. This gives us
that $\mu \sqrt{\epsilon}$ should be held fixed. Using
\eqref{gene} we find that the decoupling limit \eqref{adslimit}
translates to the following decoupling limit on the pp-wave
background \eqref{ppmet}-\eqref{ppF5}
\begin{equation}
\label{pplim}  \epsilon \rightarrow 0 \spa \mu \rightarrow \infty
\spa \tilde{\mu} \equiv \mu \sqrt{\epsilon}\ \mbox{fixed} \spa
\tilde{H}_{\rm lc} \equiv \frac{H_{\rm lc}}{\epsilon}\ \mbox{fixed}
\spa \tilde{g}_s \equiv \frac{g_s}{\epsilon}\ \mbox{fixed} \spa
l_s,\ p^+ \ \mbox{fixed}
\end{equation}
Clearly this can be seen as a large $\mu$ limit of the pp-wave.

It is important to remark that the limit \eqref{pplim} is
consistent with the Penrose limit \eqref{Plimc} since the limit
relies on having large $\tilde{R}$ and large $J$ and these are
both kept fixed in the limit \eqref{pplim}. Furthermore, we see
from \eqref{pplim} and \eqref{gene} that we have
\begin{equation}
\label{pplus} p^+ = \frac{J}{\tilde{\mu} \tilde{R}^2}
\end{equation}
so having $p^+$ fixed is consistent with having large $J$ and
large $\tilde{R}$.

We consider now the spectrum of the light-cone Hamiltonian
\eqref{fullstrspec}-\eqref{levmat} in the limit \eqref{pplim}.
First we notice that $f \rightarrow \infty$, so $f^{-1}\omega_n
\simeq 1 + n^2/(2f^2) + \CO(f^{-4})$. Therefore, most of the
excitations have $\epsilon^{-\frac{1}{2}} l_s^2 p^+  H_{\rm lc}$
of order $f$. Such excitations do not survive the limit
\eqref{pplim}. It is easy to see that this means that $N_n=0$,
$N^{(I)}=0$ and $F^{(b)}_n=0$ for $n \in \Z$. Only the excitations
connected to the number operator $M_n$ have a chance of surviving
since $\omega_n - f$ is not of order $f$ when $f \rightarrow
\infty$. Focusing on these excitations, we have
\begin{equation}
\frac{l_s^2 p^+}{\sqrt{\epsilon}} H_{\rm lc} = \sum_{n\neq 0}
(\omega_n - f) M_n \simeq \sum_{n\neq 0 } \frac{n^2}{2f} M_n
\end{equation}
We get therefore in the limit \eqref{pplim} the spectrum
\begin{equation}
\label{spec1} \tilde{H}_{\rm lc} = \frac{1}{2 \tilde{\mu} (l_s^2
p^+ )^2 } \sum_{n \neq 0} n^2 M_n \spa \sum_{n\neq 0} n M_n = 0
\end{equation}
where we also included the level matching condition obtained from
\eqref{levmat}.

We now want to show that this spectrum indeed matches the spectrum
\eqref{spectrum} obtained in weakly coupled $\CN = 4$ SYM. First we
notice that the fact that the string vacua are degenerate with
respect to the momentum $p_1$ precisely fits with the fact that the
gauge theory vacua \eqref{vacua2} are degenerate with respect to
$S_z$, as one can see explicitly from \eqref{gene}.

As a next step, we see from \eqref{pplus} and \eqref{Rtilde} that
\begin{equation}
(\tilde{\mu} l_s^2 p^+)^2 = \frac{J^2}{4\pi^2 \tilde{\lambda}}
\end{equation}
Thus, the Penrose limit \eqref{Plimc} corresponds, in terms of the
gauge theory, to the limit
\begin{equation}
\label{ppdic} \tilde{\lambda} \rightarrow \infty \spa J
\rightarrow \infty \spa \frac{\tilde{\lambda}}{J^2} \ \mbox{fixed}
\end{equation}
This fits perfectly with the fact that we want to match the
spectrum \eqref{spec1} to the spectrum of planar $\CN=4$ SYM in
the decoupling limit \eqref{su2limit} for large $\tilde{\lambda}$
and large $J=L$. Employing now \eqref{ppdic} we see that we can
rewrite \eqref{spec1} as
\begin{equation}
\label{spec2} \frac{1}{\tilde{\mu}} \tilde{H}_{\rm lc} =
\frac{2\pi^2 \tilde{\lambda}}{J^2} \sum_{n\neq 0} n^2 M_n \spa
\sum_{n\neq 0} n M_n = 0
\end{equation}
This precisely matches the spectrum \eqref{spectrum} of
$\tilde{\lambda} D_2$ on the gauge theory side, since we have
$J=L$. Notice that the $1/\tilde{\mu}$ in \eqref{spec2} origins
from \eqref{gene}, thus it is $\tilde{H}_{\rm lc}/\tilde{\mu}$ and
$\tilde{\lambda} D_2$ that one should match.

In conclusion, we have found that we can match the spectrum of
weakly coupled string theory in the pp-wave regime and in the
pp-wave decoupling limit \eqref{pplim}, with the spectrum of
weakly coupled planar $\CN=4$ SYM in the decoupling limit
\eqref{su2limit} for large $\tilde{\lambda}$ and large $J=L$. This
gives a strong indication that the induced AdS/CFT correspondence
suggested in Section \ref{sec:adslim}, between $\CN=4$ SYM in the
decoupling limit \eqref{altsu2limit} and string theory on $\ads_5
\times S^5$ in the dual decoupling limit \eqref{adslimit}, indeed
is correct.

We note that there is a geometric picture of the large $\mu$ limit
\eqref{pplim}. Since the $x^3,...,x^8$ directions have a
square-well potential with $\mu$ as coefficient, it is clear that
these directions decouple. Moreover, since only $x^1$ is a flat
direction, while the other seven transverse directions are not, it
is intuitively clear that only modes connected to the flat
direction survive. Thus, we can see on a purely geometric level
that it is the presence of a flat direction that enables us to
perform a non-trivial large $\mu$ limit in which we have finite
decoupled modes left. This is a more intuitive way to see why we
are employing the pp-wave background with a flat direction
\eqref{ppmet}-\eqref{ppF5} rather than the usual pp-wave
background used in \cite{Berenstein:2002jq} in which there are no
flat transverse directions.

Finally, we note that the limit \eqref{pplim} easily can be turned
in to a decoupling limit for a gas of strings on the pp-wave
background \eqref{ppmet}-\eqref{ppF5}, implementing the limit
\eqref{adsthermlim} on the pp-wave. This is done by supplementing
the limit \eqref{pplim} with
\begin{equation}
\label{pplimtemp} T \rightarrow 0 \spa \Omega \rightarrow 1 \spa
\epsilon = 1-\Omega \spa \tilde{T} \equiv \frac{T}{1-\Omega} \
\mbox{fixed}
\end{equation}
in accordance with the limits \eqref{adslimit} and
\eqref{adsthermlim}.

\subsection{Comments on matching of spectra}

The result of Section \ref{sec:declimpp} of the matching of the
spectra of weakly coupled gauge theory and string theory in their
respective decoupling limits is a highly non-trivial result: We
have matched the spectrum of gauge theory states in weakly coupled
gauge theory with the spectrum of free strings on a pp-wave. It is
interesting to consider how it is possible that the spectra indeed
can match. There are several underlying reasons for this:
\begin{itemize}
\item We can consider large $\tilde{\lambda}$ on the gauge theory
side even though we have $\lambda \rightarrow 0$ in the decoupling
limit \eqref{su2limit}. This ensures that only the magnon states
of the Heisenberg spin chain contribute. For $\lambda \ll 1$ with
fixed chemical potentials there would be many more states present
than the ones dual to pp-wave strings states, since this merely is
a perturbation of the spectrum of free $\CN=4$ SYM.
\item That the limit involves $E-J \rightarrow 0$ means that we
are expanding around the chiral primary states \eqref{vacua1}.
Thus, we are matching states of the gauge theory and string theory
which lie close to the chiral primaries.
\item On the gauge theory side, the Hamiltonian truncates to $H=
D_0+\tilde{\lambda}D_2$. This enables us to compute the spectrum
for large $\tilde{\lambda}$.
\item We have a pp-wave, being the pp-wave background
\eqref{ppmet}-\eqref{ppF5}, with the same vacuum structure as that
of $\CN=4$ SYM in the decoupling limit \eqref{su2limit}.
Furthermore, the pp-wave is a good approximation for large
$\tilde{\lambda}$ and $J$, which precisely is the regime that we
can match to the gauge theory side.
\item The pp-wave background \eqref{ppmet}-\eqref{ppF5} is a
maximally supersymmetric background of type IIB supergravity, and
is furthermore an $\alpha'$ exact background of type IIB string
theory (see e.g. \cite{Sadri:2003pr}). This makes the pp-wave
spectrum \eqref{fullstrspec} reliable in the decoupling limit
\eqref{pplim}.
\end{itemize}
In Section \ref{sec:strhag} we match furthermore the Hagedorn
temperature of gauge theory and string theory, in their respective
decoupling limits. That this works can be seen as a direct
consequence of the matching of the spectra.

\section{String theory Hagedorn temperature}
\label{sec:strhag}

In this section we compute the Hagedorn temperature for strings on
the pp-wave background \eqref{ppmet}-\eqref{ppF5} in the
decoupling limit \eqref{pplim}, \eqref{pplimtemp} in two different
ways. In Section \ref{sec:redhag} we compute the Hagedorn
temperature directly from the reduced pp-wave spectrum
\eqref{spec1}. In Section \ref{sec:haglim} we instead take the
decoupling limit \eqref{pplim}, \eqref{pplimtemp} of the Hagedorn
temperature for the full pp-wave spectrum \eqref{fullstrspec}.
Both of these computations give the same result, which we show can
be matched with the Hagedorn temperature \eqref{haggauge} computed
in weakly coupled $\CN=4$ SYM.

\subsection{Hagedorn temperature for reduced pp-wave spectrum}
\label{sec:redhag}

In this section we compute the Hagedorn temperature for the
reduced pp-wave spectrum \eqref{spec1}. This is the spectrum
obtained for type IIB superstring theory in the pp-wave background
\eqref{ppmet}-\eqref{ppF5} in the decoupling limit \eqref{pplim}.
We show that the result for the Hagedorn temperature coincides
with the one of the dual gauge theory \eqref{haggauge}.

We consider first the multi-string partition function
\begin{equation}
\log Z (\tilde{a},\tilde{b},\tilde{\mu}) = \sum_{n=1}^{\infty}
\frac{1}{n} \mbox{Tr} \left( e^{-\tilde{a} n \tilde H_{{\rm
l.c.}}-\tilde{b}np^+}\right) \label{fab}
\end{equation}
where the trace is taken over single-string states with spectrum
\eqref{spec1}. The parameters $\tilde{a}$ and $\tilde{b}$ can be
viewed as inverse temperature and chemical potential,
respectively, for the pp-wave strings. We find the values for
$\tilde{a}$ and $\tilde{b}$ in terms of the $\ads_5 \times S^5$
parameters below. Note that we do not have fermions in the
spectrum. The measure for the trace over $p^+$ is
$\frac{l}{2\pi}\int_0^{\infty} dp^+$, where $l$ is the (infinite)
length of the $9$'th dimension. We get
\begin{equation}
\log Z = - \sum_{n=1}^{\infty}\frac{l\tilde{\beta}}{8\pi^2{l}_s^2}
\int_{0}^{\infty} \frac{d\tau_2}{\tau_2^2}
\int_{-\frac{1}{2}}^{\frac{1}{2}}d\tau_{1}\sum_{M_{{m}}=0}^{\infty}
e^{-\frac{\tilde{b}n^2\tilde{\beta}}{4\pi{l}_s^2\tau_2}-\frac{2\pi\tilde{a}
\tau_2}{\tilde{\beta}\tilde f}\sum_{m\neq 0}m^2M_m+2\pi i
\tau_{1}\sum_{m\neq 0}mM_{m}}
\end{equation}
where the level matching condition is imposed by introducing an
integration over the Lagrange multiplier $\tau_1$ and we
introduced the quantities
\begin{equation}
\tau_{2}=\frac{n\tilde{\beta}}{4\pi{l}_s^2 p^{+}}~,~~~ \tilde
f={l}_s^2 p^{+}\tilde{\mu}=\frac{n\tilde{\beta}
\tilde{\mu}}{4\pi\tau_{2}} \label{mu}
\end{equation}
Summing over the occupation number we get
\begin{equation}
\log Z = - \sum_{n=1}^{\infty}\frac{l\tilde{\beta}}{8\pi^2{l}_s^2}
\int_{0}^{\infty} \frac{d\tau_2}{\tau_2^2}
\int_{-\frac{1}{2}}^{\frac{1}{2}}d\tau_{1}
e^{-\frac{\tilde{b}n^2\tilde{\beta}}{4\pi{l}_s^2\tau_2}}
|G(\tau_1,\tau_2,\tilde f)|^2 \label{freemu}
\end{equation}
where the generating function $G$ is given by
\begin{equation}
G(\tau_1,\tau_2,\tilde f)=\prod_{m=1}^{\infty}
\left(\frac{1}{1-e^{-\frac{2\pi\tilde{a}\tau_2}{\tilde{\beta}\tilde{f}}
m^2+2\pi i\tau_1 m}}\right)
\end{equation}
To see where the partition function diverges we need to estimate
the asymptotic behavior of the function $G$. This is done in
Appendix \ref{app:Asy} were we show that it diverges in the limit
$\tau_2\to 0$. More precisely, one can show that for $\tau_2$ that
goes to zero, there is a divergence only if $\tau_1=0$ and the
leading contribution is given by
\begin{equation}
G(0,\tau_2,\tilde f)\sim
\exp\left(\zeth\sqrt{\frac{\tilde{\beta}\tilde
f}{8\tilde{a}\tau_2}}\,\right)=
\exp\left(\zeth\frac{\tilde{\beta}}{4\tau_2}\sqrt{\frac{n\tilde{\mu}}{2\pi
\tilde{a}}}\,\right) \label{gasy}
\end{equation}
After substituting this result in the expression for the partition
function \eqref{freemu} in the limit $\tau_2\to 0$ we find that
we have a Hagedorn singularity for%
\footnote{We note that to gain a better understanding of the
behavior of the partition function \eqref{freemu} one should
perform the integral over $\tau_1$. This however would just
produce a different power of $\tau_2$ in the prefactor of the
partition function and it would not modify the result
\eqref{hageab} for the Hagedorn temperature.}
\begin{equation}
\tilde{b} \sqrt{\tilde{a}} =  l_s^2 \zeth \sqrt{2\pi\tilde{\mu}}
\label{hageab}
\end{equation}
where the relevant contribution is given by the $n=1$ mode.

In order to compare \eqref{hageab} with the gauge theory result
\eqref{haggauge} we have to express the parameters $\tilde{a}$ and
$\tilde{b}$ in terms of the gauge theory
quantities~\cite{Brower:2002zx}. Using Eqs.~\eqref{gene} and
\eqref{inddic} it is not difficult to see that $\tilde{a}$ and
$\tilde{b}$ should be identified in the following way
\begin{equation}
\tilde{a}=\frac{\tilde{\beta}}{\tilde{\mu}} \spa \tilde{b}=4\pi
l_s^2 \tilde{\beta}\tilde{\mu} \tilde{T}_{\rm str}= 2\pi l_s^2
\tilde{\beta}\tilde{\mu}\sqrt{\tilde{\lambda}} \label{ab}
\end{equation}
With these identifications Eq.~\eqref{hageab} gives
\begin{equation}
\tilde{T}_H = (8\pi)^{\frac{1}{3}} \left[ \zeth
\right]^{-\frac{2}{3} } \tilde{T}_{\rm str}^{\frac{2}{3}}  =
(2\pi)^{\frac{1}{3}} \left[ \zeth \right]^{-\frac{2}{3} }
\tilde{\lambda}^{\frac{1}{3}} \ \ \ \label{hagstring}
\end{equation}
which precisely coincides with the result \eqref{haggauge}
obtained on the gauge theory side.

We have thus shown that the Hagedorn temperature of type IIB
string theory on $\ads_5 \times S^5$ in the decoupling limit
\eqref{adsthermlim} matches with the Hagedorn/deconfinement
temperature \eqref{haggauge} computed in weakly coupled $\CN=4$
SYM in the dual decoupling limit \eqref{su2limit}. This is done in
the regime of large $\tilde{\lambda}$. On the string side we
obtained the Hagedorn temperature by considering the large
$\tilde{\lambda}$ and $J$ limit corresponding to strings on the
pp-wave background \eqref{ppmet}-\eqref{ppF5} in the decoupling
limit \eqref{pplim}. The result means that in the sector of
AdS/CFT defined by the decoupling limits \eqref{adsthermlim} and
\eqref{su2limit} we can indeed show that the Hagedorn temperature
for type IIB string theory on the $\ads_5\times S^5$ background is
mapped to the Hagedorn/deconfinement temperature of weakly coupled
planar $\CN=4$ SYM on $\R \times S^3$. Thus we have direct
evidence that the confinement/deconfinement transition found in
weakly coupled planar $\CN=4$ SYM on $\R \times S^3$ is linked to
a Hagedorn transition of string theory on $\ads_5 \times S^5$, as
conjectured in
\cite{Witten:1998zw,Sundborg:1999ue,Polyakov:2001af,Aharony:2003sx}.

Note that the matching of the Hagedorn temperature made above to
some extent follows directly from the matching of the spectra made
in Section \ref{sec:conpp}. However, to check that the computation
of the Hagedorn temperature indeed is consistent with taking the
decoupling limit \eqref{pplim}, \eqref{pplimtemp} of strings on
the pp-wave background \eqref{ppmet}-\eqref{ppF5} we check in the
following section that one can find the same Hagedorn temperature
directly by taking the decoupling limit on the Hagedorn
singularity for the full pp-wave spectrum \eqref{fullstrspec}.

\subsection{Limit of Hagedorn temperature for full pp-wave
spectrum} \label{sec:haglim}

In this section we show that by computing the Hagedorn temperature
using the full spectrum \eqref{fullstrspec} and subsequently
taking the limit \eqref{pplim}, \eqref{pplimtemp} we obtain again
the result \eqref{hagstring} for the Hagedorn temperature.

We consider the multi-string partition function
\begin{equation}
\log Z(a,b,\mu)=\sum_{n=1}^{\infty}\frac{1}{n}\mbox{Tr}\left(
(-1)^{(n+1) {\rm \bf F}} e^{-a n H_{{\rm
l.c.}}-bnp^+}\right)\label{fab2}
\end{equation}
where the trace is over single-string states with the spectrum
\eqref{fullstrspec}, and ${\rm \bf F}$ is the space-time fermion
number. The computation of the partition function \eqref{fab2} is
similar to that of the reduced spectrum done in Section
\ref{sec:redhag} and it has been done in
Ref.~\cite{Sugawara:2003qc} for $b=0$.%
\footnote{In~\cite{Sugawara:2003qc} the direction $x^1$ is
compactified and it is shown that only the sector with zero
winding number contributes to the partition function.}
Generalizing the computation to non-zero $b$, we get that the
Hagedorn singularity occurs for
\begin{equation}
b= 4 l_s^2 \mu \sum_{p=1}^{\infty}\frac{1}{p}\left[3 + \cosh(\mu a
p) - 4(-1)^p\cosh \! \left(\frac{1}{2} \mu ap \right) \right] K_1(
\mu a p ) \label{hagabeq}
\end{equation}
where $K_\nu(x)$ is the modified Bessel function of the second
kind. Using \eqref{gene} we see that we should identify
\begin{equation}
a=\frac{\mu \tilde{\beta}}{\tilde{\mu}^2} \spa b = 4\pi \mu l_s^2
T_{\rm str} \tilde{\beta} \label{ab2}
\end{equation}
We now take the limit \eqref{pplim}, \eqref{pplimtemp}. The Bessel
function can be approximated by its behavior for large values of
the argument
\begin{equation}
K_1(x)\sim e^{-x}\sqrt{\frac{\pi}{2}}\left(\sqrt{\frac{1}{x}}
+{\cal{O}}(x^{-3/2})\right)
\end{equation}
It is easy to see that in this limit only the $\frac{1}{2}e^{\mu a
p}$ term inside the $[\cdots]$ paranthesis in \eqref{hagabeq}
survives. We note that this is precisely the contribution from the
$M_n$ oscillators in \eqref{fullstrspec}. To see that the other
terms in \eqref{hagabeq} vanish it is enough to consider $p=1$
since the higher $p$ terms are exponentially suppressed. From the
surviving term it is then straightforward to show that we again
get the Hagedorn temperature \eqref{hagstring}, which matches the
gauge theory result \eqref{haggauge}.

We can conclude from the above that taking the decoupling limit
\eqref{pplim}, \eqref{pplimtemp} on the spectrum
\eqref{fullstrspec} on the pp-wave \eqref{ppmet}-\eqref{ppF5} is
consistent with taking the decoupling limit of the Hagedorn
singularity on the pp-wave. I.e. taking the decoupling limit
before computing the Hagedorn temperature commutes with computing
the Hagedorn temperature and then subsequently taking the
decoupling limit. This is a good check on the consistency of the
decoupling limit \eqref{pplim}, \eqref{pplimtemp}.

\section{Discussion and conclusions}
\label{sec:concl}

\begin{figure}[ht]
\centerline{\epsfig{file=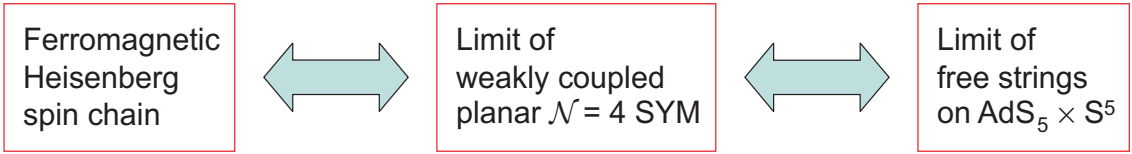,width=15cm,height=2cm} }
\caption{A spin chain/gauge theory/string theory triality.
\label{fig:trial}}
\end{figure}

The general idea of this paper is that by taking a certain
decoupling limit we get a self-consistent decoupled sector of the
AdS/CFT correspondence. On the gauge theory side, we take the
decoupling limit \eqref{su2limit} of $SU(N)$ $\CN=4$ SYM on $\R
\times S^3$. On the string theory side, we take the decoupling
limit \eqref{adslimit} (see also \eqref{adsthermlim}) of type IIB
strings on $\ads_5\times S^5$. In \cite{Harmark:2006di} it was
shown that the sector of planar $\CN=4$ SYM on $\R \times S^3$
obtained in the decoupling limit \eqref{su2limit} also is
described by the ferromagnetic Heisenberg spin chain, as reviewed
in Section \ref{sec:declim}. On the string theory side, the planar
limit of $\CN=4$ SYM corresponds to free strings propagating on
$\ads_5\times S^5$. We have thus the spin chain/gauge
theory/string theory triality depicted in Fig.~\ref{fig:trial}.
Since the Heisenberg chain is integrable, we get that both the
gauge theory and the string theory should be integrable. In this
sense we have found a solvable sector of AdS/CFT. One of the
important features of the triality of Fig.~\ref{fig:trial} is that
we are considering small 't Hooft coupling $\lambda \rightarrow 0$
on the gauge theory side. On the string theory side this
corresponds to having a small string tension $T_{\rm str}$.

We have succeeded in this paper to show that the low energy
spectrum \eqref{spectrum} obtained on the spin chain/gauge theory
side matches the spectrum of free strings on a maximally
supersymmetric pp-wave background. With this, we have shown that
the low energy part of the spectrum of the gauge theory and string
theory sides of the triality of Fig.~\ref{fig:trial} matches. This
is a rather non-trivial result in that we have obtained a string
theory spectrum, which is calculable on the string theory side,
directly in weakly coupled gauge theory. Indeed, to our knowledge,
this is the first non-trivial matching in AdS/CFT done between
gauge theory and string theory in the $\lambda \ll 1$ regime.

Related to this result, we have shown that the
Hagedorn/deconfinement temperature in weakly coupled planar
$\CN=4$ SYM on $\R \times S^3$ in the limit \eqref{su2limit}
matches the Hagedorn temperature of weakly coupled string theory
on a maximally supersymmetric pp-wave background
\eqref{ppmet}-\eqref{ppF5} in the decoupling limit \eqref{pplim},
\eqref{pplimtemp}. This shows that the confinement/deconfinement
transition found in weakly coupled planar $\CN=4$ SYM on $\R
\times S^3$ is linked to a Hagedorn transition of string theory on
$\ads_5 \times S^5$, as conjectured in
\cite{Witten:1998zw,Sundborg:1999ue,Polyakov:2001af,Aharony:2003sx}.

The mechanism behind these successful matches between string
theory and gauge theory is the $SU(2)$ decoupling limit found in
\cite{Harmark:2006di}. In this decoupling limit we consider the
gauge theory states lying very close to a certain chiral primary
sector (defined by $D = J$). This enables us to decouple most of
the gauge theory states leaving only the $SU(2)$ sector, and the
Hamiltonian truncates to \eqref{Hsu2}, which has the consequence
that we can study the decoupled sector for finite
$\tilde{\lambda}$. On the string theory side, we find that the
Penrose limit \cite{Bertolini:2002nr} of $\ads_5\times S^5$
leading to the pp-wave background \eqref{ppmet}-\eqref{ppF5} with
a flat direction gives a pp-wave string spectrum for which the
vacua precisely are dual to the chiral primary states expanded
around on the gauge theory side. Translating the dual decoupling
limit for string on $\ads_5\times S^5$ into a decoupling limit for
the pp-wave enables us to study the decoupled sector from the
string theory side. Unlike the usual gauge-theory/pp-wave
correspondence we can match the gauge theory and string theory
spectra for small 't Hooft coupling $\lambda \rightarrow 0$ since
for finite $\tilde{\lambda}$ only the gauge theory states in the
$SU(2)$ sector close to the chiral primary states contribute at
low energies.

\subsubsection*{Future directions}

One of the most interesting extensions of the matching of the
Hagedorn temperature between gauge theory and string theory of
this paper would be to reproduce the $\tilde{\lambda}^{-1/3}$
correction from string theory side. From the thermodynamics of the
Heisenberg chain, we found the correction \eqref{haggauge2}. On
the string theory side, computing this correction would involve
going away from the large $J$ limit. More generally, it would be
highly interesting to match finite size corrections to the
spectrum of the Heisenberg chain, to $1/J$ corrections to the
pp-wave spectrum.

Another interesting class of corrections to consider would be to
look at corrections coming from terms of order
$\tilde{\lambda}\lambda$ in the Hamiltonian. I.e. in
\cite{Harmark:2006di} we have that the leading correction for
small $\lambda$ to the Hamiltonian for the $SU(2)$ sector is
\begin{equation}
H = D_0 + \tilde{\lambda} D_2 + \tilde{\lambda}\lambda D_4 + \CO(
\tilde{\lambda}\lambda^2 )
\end{equation}
In this regime one could be worried about corrections coming from
the fact that states outside the $SU(2)$ sector are not completely
decoupled. However, we do not expect that to be important, since
such corrections appear non-perturbatively in terms of the expansion
parameter $1-\Omega$ \cite{Harmark:2006di}.

Considering $\lambda$ corrections could be very important for a
better understanding of the three-loop discrepancy
\cite{Beisert:2004hm,Callan:2004uv,Callan:2004ev} between
anomalous dimensions computed in $\CN=4$ SYM and string energies
for strings on $\ads_5\times S^5$. The reason for the three-loop
discrepancy could very well be that there are interpolating
functions in $\lambda$ that one does not see when doing a naive
large $\lambda$ extrapolation of the gauge theory results. For our
decoupled sector we do not have any need for interpolating
functions, since we are not extrapolating the anomalous dimensions
to infinite $\lambda$. Therefore, it would be rather interesting
in this light to see if there is a discrepancy for $\lambda$
corrections to our decoupled sector.

One could furthermore consider other decoupling limits. In
\cite{Harmark:2006di} we found a decoupling limit of planar
$\CN=4$ SYM on $\R \times S^3$ in which it decouples to the
$SU(2|3)$ spin chain, in a very similar way as that of the $SU(2)$
decoupling limit considered in this paper. We expect similar
results for this sector. This could be interesting to work out
since the spectrum is more complicated due to the presence of
fermions. As mentioned in \cite{Harmark:2006di} it is moreover
conceivable that there are other interesting decoupling limits of
supersymmetric gauge theories with less supersymmetry, hence one
could hope to match the spectrum and Hagedorn temperatures for
such cases as well. In particular, it would be interesting to
consider generalizing the $SU(2)$ decoupling limit of
\cite{Harmark:2006di} used in this paper to $\CN=2$ quiver gauge
theories dual to the pp-wave background \eqref{ppmet}-\eqref{ppF5}
with $x^1$ compactified, following \cite{Bertolini:2002nr}.

Finally, we note it would be very interesting to consider
non-planar corrections to the partition function on the gauge
theory side. In \cite{Harmark:2006di} the decoupling limit also
works for finite $N$, thus it should be possible to gain more
information about the Hagedorn/deconfinement phase transition, for
example whether it is a first order phase transition or not.%
\footnote{In this connection one could also hope to get a better
understanding of the small black hole in $\ads_5 \times S^5$ from
the gauge theory point of view
\cite{Alvarez-Gaume:2005fv,Alvarez-Gaume:2006jg,Hollowood:2006xb}.}

\section*{Acknowledgments}

We thank Gianluca Grignani for many nice discussions and useful
suggestions. We thank Jaume Gomis and Peter Orland for useful
discussions. We thank Konstantin Zarembo for useful discussions
and in particular for pointing out to us the paper
\cite{PhysRevLett.89.117201}. The work of M.O. is supported in
part by the European Community's Human Potential Programme under
contract MRTN-CT-2004-005104 `Constituents, fundamental forces and
symmetries of the universe'.

\begin{appendix}

\section{Asymptotic behavior of the generating function}
\label{app:Asy}

In this appendix we will show how to estimate the asymptotic
behavior of the function
\begin{equation}
G(a,b)=\prod_{n=1}^{\infty}\frac{1}{1-e^{-an^2+ibn}}
\label{genfunc}
\end{equation}
with $a$ and $b$ real and $a>0$. The previous expression can be
written as
\begin{equation}
G(a,b)=\exp\left\{\sum_{n=1}^{\infty}\sum_{p=1}^{\infty}\frac{e^{-apn^2+ibpn}}{p}\right\}
\label{gf}
\end{equation}
We are interested in studying the $a\to 0$ limit.

Consider first the case $b\neq 0$. In the limit $a\to 0$ the sum
over $n$ in~\eqref{gf} can be replaced by an integral and we have
\begin{equation}
G(a,b)\sim\exp\left\{\sum_{p=1}^{\infty}\int_{1}^{\infty}dx\frac{e^{-apx^2+ibpx}}{p}\right\}
=\exp\left\{\sum_{p=1}^{\infty}\sqrt{\frac{\pi}{a}}\frac{e^{-b^2p/4a}}{2p}
{\rm
Erfc}\left(\sqrt{pa}-i\frac{b\sqrt{p}}{2\sqrt{a}}\right)\right\}\label{beh}
\end{equation}
where ${\rm Erfc}(x)$ is the complementary error function (${\rm
Erfc}(x)=1-{\rm erf}(x)$ where ${\rm erf}(x)$ is the error
function). For $a\to 0$ and $b\neq 0$ the complementary error
function can be approximated as
\begin{equation}
{\rm Erfc}\left(\sqrt{pa}-i\frac{b\sqrt{p}}{2\sqrt{a}}\right)\sim
2 i\sqrt{\frac{a}{\pi p b^2}}e^{b^2 p/4a}
\end{equation}
so that the generating function becomes
\begin{equation}
G(a,b)\sim \exp\left\{\frac{i}{b}\zeth\right\}
\end{equation}
where $\zeta(x)$ is the Riemann zeta function. We thus see that
for $b\neq 0$ there is no divergent contribution.

To extract the divergent contribution we set $b=0$
in~\eqref{genfunc} so that
\begin{equation}
G(a,0)=\prod_{n=1}^{\infty}\frac{1}{1-e^{-an^2}}\sim
\exp\left[F(a)\right]\label{genfunc2}
\end{equation}
where we defined
\begin{equation}
\label{defFa} F(a) \equiv
-\int_{1}^{\infty}dx\log\left(1-e^{-ax^2}\right)
\end{equation}
Here we have again approximated the sum over $n$ by an integral.
Introducing the new variable $y=x\sqrt a$ we have that
\begin{equation}
\lim_{a\to
0}\sqrt{a}\,F(a)=-\int_0^{\infty}dy\log\left(1-e^{-y^2}\right)=
\sum_{p=1}^{\infty}\int_0^{\infty}dy\frac{e^{-y^2p}}{p} =
\frac{\sqrt{\pi}}{2}\zeth
\end{equation}
Thus, we see from this that for $b=0$ there is a divergent
contribution in \eqref{genfunc} in the $a\to 0$ limit, giving
\begin{equation}
\label{Ga0} G(a,0)\sim\exp\left\{ \zeth
\sqrt{\frac{\pi}{4a}}\,\right\}
\end{equation}
This is the leading asymptotic behavior of $G(a,0)$ for
$a\rightarrow 0$.

\end{appendix}



\providecommand{\href}[2]{#2}\begingroup\raggedright\endgroup

\end{document}